\documentclass{article} % For LaTeX2e
\usepackage{iclr2025_conference,times}

\usepackage{hyperref}
\usepackage{url}
\usepackage{siunitx}
\usepackage{multirow}
\usepackage{booktabs}
\usepackage{graphicx}
\usepackage{subcaption}
\usepackage{float}
\usepackage{tablefootnote}
\usepackage{makecell}
\usepackage{authblk}
\usepackage{amsmath,amsfonts,bm}

\title{Flow-Based Fragment Identification via Binding Site-Specific Latent Representations}

\newcommand{\ts}{\textsuperscript}

\author[1]{Rebecca M.~Neeser}
\author[1]{Ilia Igashov}
\author[1]{Arne Schneuing}
\author[1,2,3,4]{\authorcr Micheal Bronstein} 
\author[1,5]{Philippe Schwaller}
\author[1]{Bruno Correia}
\affil[1]{Ecole Polytechnique Fédérale de Lausanne, Switzerland}
\affil[2]{Oxford University, UK}
\affil[3]{VantAI, USA}
\affil[4]{Aithyra Research Institute for Biomedical Artificial Intelligence, Austria}
\affil[5]{National Centre of Competence in Research (NCCR) Catalysis EPFL, Switzerland }
\affil[ ]{\texttt{bruno.correia@epfl.ch}}

\iclrfinalcopy 
\begin{document}

\maketitle

\begin{abstract}
% motivation
Fragment-based drug design is a promising strategy leveraging the binding of small chemical moieties that can efficiently guide drug discovery. 
% problem statement
The initial step of fragment identification remains challenging, as fragments often bind weakly and non-specifically.
% approach
We developed a protein-fragment encoder that relies on a contrastive learning approach to map both molecular fragments and protein surfaces in a shared latent space. The encoder captures interaction-relevant features and allows to perform virtual screening as well as generative design with our new method LatentFrag. In LatentFrag, fragment embeddings and positions are generated conditioned on the protein surface while being chemically realistic by construction.
% results
Our expressive fragment and protein representations allow location of protein-fragment interaction sites with high sensitivity and we observe state-of-the-art fragment recovery rates when sampling from the learned distribution of latent fragment embeddings. Our generative method outperforms common methods such as virtual screening at a fraction of its computational cost providing a valuable starting point for fragment hit discovery. We further show the practical utility of LatentFrag and extend the workflow to full ligand design tasks.
% conclusion
Together, these approaches contribute to advancing fragment identification and provide valuable tools for fragment-based drug discovery.
\end{abstract}

%%%%%%%%%%%%%%%%%%%%%%%%%%%%%%%%%%%%%%%%%%%%%%%%%%%%%%%

\section{Introduction}
Hit identification remains a critical challenge in drug discovery, despite advances in screening techniques and computational tools~\citep{jalencas2024design, hasselgren2024artificial}. Traditional high-throughput screening~(HTS) workflows to identify ligands targeting  proteins of interest have limitations, particularly in exploring efficiently large chemical spaces and identifying actionable starting points for drug design~\citep{edfeldt2011fragment}. Fragment-Based Drug Design (FBDD) offers
a promising alternative, leveraging smaller chemical moieties that can be combined to form potent ligands~\citep{congreve2008recent, hubbard2011experiences}.\par 

Fragment screening has several advantages over conventional HTS approaches, that use larger, drug-like molecules. Smaller sized fragments typically exhibit higher ligand efficiency, a measure of binding energy per atom, and their smaller size allows exploration of bigger chemical space~\citep{congreve2008recent}. Although fragments generally have very low binding affinities than larger ligands, the combination of multiple key interactions with the target protein can yield a more specific ligand~\citep{edfeldt2011fragment, yu2020general}. \par

Recent machine learning~(ML) approaches enable rapid exploration of chemical space, yet many FBDD methods do not account for a protein structure, rely on prior knowledge of fragment hits~\citep{mccorkindale2022fragment}, or depend on classical molecular docking~\citep{bian2018computational, SEED}.
ML-based Structure-based Drug Design~(SBDD) incorporates three-dimensional information but frequently generates unrealistic or synthetically inaccessible molecules when designing full ligands in an all-atom approach~\citep{buttenschoen2024posebusters, yang20243d}. This challenge is less pronounced in FBDD as the use of known fragments constrains the chemically accessible space. However, despite \textit{in silico} methods trying to tackle this problem~\citep{imrie2020deep,neeser2023reinforcement,igashov2024equivariant,ferla2025fragmenstein}, challenges of merging, linking and growing fragments to a full ligand remain. Fragment docking often is unable to localise binding sites within protein pockets as fragments often lack the necessary structural information content for docking algorithms to identify good binding poses and meaningful docking scores. These challenges further emphasise the need for new, more effective, structure-based FBDD solutions~\citep{de2021fragment}.

To address these limitations, we introduce a novel structure-based fragment screening approach that learns a protein-fragment representation through contrastive training. Our encoder maps protein surfaces and molecular fragments into a shared latent space. These rich fragment representations capture aspects of binding interactions with the target while maintaining chemical relevance through fragment similarity-based penalties. The resulting latent representations can be directly used in virtual screening. Inspired by \citet{igashov2022decoding}, who performed ligand and fragment screening based on pocket representations but using a model trained in protein-protein interactions, our model is specifically tailored for fragment screening and does not depend on the availability of fragments in crystal structures.

Beyond this, we extend the method to a generative framework for structure-based fragment identification called LatentFrag. Instead of relying on a fixed fragment library, our generative approach, explicitly conditioned on the protein pocket, learns distributions of fragments and their spatial organization directly in the latent space. This approach does not require a decoder, but rather queries a fragment library for the most similar fragment embedding ensuring chemically realistic fragments. This mechanism allows us to replace or expand the fragment library without retraining the generative model or even without re-sampling once embeddings were generated for one target. By representing ligands as fragment graphs during training, LatentFrag is able to generate more than one fragment per model call increasing efficiency. To train our models, we curated a dataset of pairs of proteins and fragmented drug-like ligands from the Protein Data Bank~(PDB)~\citep{berman2000protein}. The fragmentation was carried out in a manner that yields synthetically sensible fragments while ensuring sufficient binding specificity by limiting the minimal size~\citep{degen2008art}. 

By assessing both “hard” recovery metrics and “soft” pharmacophoric similarity, we provide insights into the model’s ability to identify meaningful fragment hits. Our analysis demonstrates improvements over virtual screening baselines, like docking, while improving computational efficiency. We further demonstrate in a case study how our framework can be complemented with existing methods~\citep{igashov2024equivariant} to extend our fragment hits to full drug-like ligands. We showcase this pipeline by targeting the therapeutically relevant protein c-Met finding many fragments recovering known interactions but also many of which have new interactions to the target. The most promising fragments are subsequently connected resulting in a structure with improved \textit{in silico} properties over the reference ligand BMS-777607.

% To summarize, our contributions are the following:
% \begin{itemize}
%     \item \textbf{Novel Protein-Fragment Contrastive Learning Approach:} We introduce a protein-fragment encoder trained in a contrastive fashion that jointly learns rich representations of protein surfaces and molecular fragments in a shared latent space. Our approach captures aspects of interaction while maintaining chemical relevance through fragment similarity-based penalties.
%     \item \textbf{Generative Structure-Based Fragment Identification:} We propose LatentFrag, a flow matching framework for fragment identification that operates directly in the protein-fragment latent space. Unlike existing methods, our approach is explicitly conditioned on protein structure and guarantees chemically valid and likely synthesizable outputs through library-based sampling.
%     \item \textbf{Evaluation Framework:} By assessing both “hard” recovery metrics and “soft” pharmacophoric similarity, we provide insights into the model’s ability to identify meaningful fragment hits. Our analysis demonstrates improvements over virtual screening baselines, like docking, while improving computational efficiency. 
% \end{itemize}

%%%%%%%%%%%%%%%%%%%%%%%%%%%%%%%%%%%%%%%%%%%%%%%%%%%%%%%

\section{Previous Work}
\label{sec:previous}
% representation learning
Molecular representation learning~(MRL) is a well-established field, with numerous studies refining and adapting methods for specific tasks such as property and reaction prediction~\citep{guo2022graph}. Various MRL approaches leverage language models as encoders~\citep{shin2019self,chithrananda2020chemberta,li2021mol} or introduce specialised representations, such as UniMol, which incorporates 3D conformers~\citep{zhouuni}, and MolR, which is tailored for reaction-based learning~\citep{wang2021chemical}. However, these methods typically do not explicitly account for protein targets and are not primarily designed for hit screening or drug discovery. \citet{gao2023drugclip} proposed DrugCLIP, which contrastively learns pocket and ligand representations for the task of virtual screening based on the UniMol encoder architecture~\citep{zhouuni}. This method conceptually shares many aspects with our work but encodes the pocket globally and full ligands making the task of fragment placement impracticable. While none of the aforementioned approaches focus specifically on molecular fragments, \citet{chakravarti2018distributed} propose a fragment-based method, though it remains centred on chemical properties relevant to tasks like property prediction rather than interaction-driven applications. A concurrent work by \citet{lohmann2024protein} proposes a conceptually similar encoder of both protein and full ligand in 3D, which allows to analyse protein pockets in latent space. However, the representation is obtained by training for the task of affinity prediction and based on the protein graph instead of the surface.\par 

% fragment identification
The task of computational fragment identification has been mostly dependent on fragment docking~\citep{bian2018computational, SEED}, which is less accurate than ligand docking. FRESCO~\citep{mccorkindale2022fragment} is a ML-based method that implicitly considers target structure through pharmacophore distributions. The extraction of those, however, requires known hits from fragment screens and respective crystal structures, which is often not available in a drug discovery campaign. The closest approach to ours by \citet{igashov2022decoding} matches protein pocket embeddings in order to find related fragment hits, making it also limited by the availability of crystal structures.

%%%%%%%%%%%%%%%%%%%%%%%%%%%%%%%%%%%%%%%%%%%%%%%%%%%%%%%

\section{Methods}
\label{sec:methods}

\subsection{Protein-Fragment Contrastive Learning}
\label{sec:methods:encoder}
The protein-fragment encoder is trained contrastively and is designed to produce expressive latent embeddings for both fragments and protein surfaces (for details see Appendix~\ref{si:encoder}). Thus, the latent vectors capture critical features for binding interactions and are uniquely suited to the task of fragment identification.

Training involves maximizing the cosine similarity between embeddings of fragments and nearby surface points on the protein (positive examples), while minimizing the similarity for other surface points elsewhere on the protein (negative examples), ensuring a robust distinction. Negative examples are selected to include both convex and concave protein surface geometries from the pocket. This prevents overfitting to concave regions, abundant among positive examples. The surface curvature is predicted on the fly by a concurrently trained classifier. Protein surface embeddings are parametrised by a geodesic convolutional neural network similar to dMaSIF~\citep{dmasif} while fragments are processed by a graph transformer~\citep{dwivedi2020generalization,vignac2022digress}. Further details are provided in Appendix~\ref{si:enc_architecture}.

To ensure chemical relevance, a fragment similarity penalty~(FSP) is incorporated via a hinge loss. This loss discourages molecules with low Tanimoto similarity from having similar embeddings, as shown in Figure~\ref{fig:enc}A. In order to integrate interaction-specific information we employ an additional classification loss. An additional classification loss is used to train the model to predict the type of non-covalent interaction~(NCI), if present, that each protein surface point can engage in. In this way, interaction-specific information is integrated.

Notably, the relative positions of the molecules are only used for assigning positive and negative examples during training, and do not directly influence the embeddings as fragments are represented as 2D graphs. This design promotes flexible, position and conformation-agnostic representations of fragment and surface features, enabling their broad applicability in different FBDD scenarios.

\subsection{LatentFrag: Fragment Identification via Flow Matching}
\label{sec:methods:FM}
To identify and place relevant fragments in a given protein pocket, we introduce LatentFrag. LatentFrag is a generative modelling approach using flow matching~\citep{lipman2022flow}, representing proteins as surface point clouds and ligands as coarse fragment graphs. Protein surface points are featurized by latent vectors and ligand nodes representing fragments are defined by a latent embedding (fragment type) and the arithmetic mean of their coordinates (centroid). The latent vectors are learned embeddings from our protein-fragment encoder. A schematic overview of the fragment identification process is shown in Figure~\ref{fig:fm}A and the neural network is detailed in Figure~\ref{si:fig:GVP_scheme}.

LatentFrag learns to map noise to structured data by matching modelled probability flows. This enables efficient sampling from complex distributions. Two flows are employed: a spherical flow for the latent fragment embeddings, which assumes a unit sphere prior, and a Euclidean flow for the centroids, with a Gaussian prior. Once the latent fragment representations have been generated with LatentFrag, we query the precomputed library by cosine similarity and place the most similar fragments. This approach ensures fragment consistency with respect to chemical plausibility and geometry. We use a library with 41,224~unique fragments extracted from the PDB~\citep{berman2000protein} (Appendix~\ref{si:data}).

The predicted coordinate represents the centre of the fragment, and obtaining its orientation requires downstream docking processes. This framework offers flexibility while aligning with fragment-based drug design workflows where a successful campaign is dependent on robust fragment identification. Detailed information on the generative frameworks is described in the Appendix~\ref{si:fm}.

\subsection{Evaluation of Fragment Identification Task}

We evaluated fragment identification through two approaches: virtual screening~(VS) using our latent embeddings (Latent VS) and the generative framework for sampling fragment embeddings and their centroids. These are compared against virtual screening based on docking (Docking VS) and a random baseline, evaluated on 100 protein targets with $\le30\%$ sequence similarity to the training set.

For Latent VS, fragments are ranked by the sum of cosine similarities between the fragment embeddings and the protein pocket surface points. The top 100 fragments per target are then selected for evaluation. Using LatentFrag, for each surface 100 samples are generated with number of fragments corresponding to the reference number of fragments. Generated latent representations are used to query a library for the closest fragments based on cosine similarity. Sampled fragments are subsequently docked using Gnina~\citep{mcnutt2021gnina} within a restricted volume around predicted centroids. We evaluate using three key metrics:
\begin{itemize}
    \item \textbf{Hard Recovery - Sampling Hits:} Total number of generated fragments exactly matching references
    \item \textbf{Hard Recovery - Unique Fragments:} Number of unique reference fragments recovered
    \item \textbf{Soft Recovery:} Shape and pharmacophoric similarity to the full reference ligand via SuCOS score~\citep{leung2019sucos}
\end{itemize}
The reference fragment count for the calculation of recovery rates often exceeds generated fragments per target and sample due to data augmentation combining BRICS fragmentation~\citep{degen2008art} with graph partitioning, allowing reference fragments to be substructures of others. Detailed information can be found in Appendix~\ref{si:eval_ident}.

\subsection{Data}
\label{sec:data}
To build the fragment library, we extracted protein-ligand structures from the PDB~\citep{berman2000protein} and remove ligands irrelevant to the task such as solvents and buffers. Subsequently ligands are fragmented using BRICS rules~\citep{degen2008art} while not allowing double bonds to be broken and reassemble fragments to reach a minimum fragment size of 8~heavy atoms in a combinatorial manner. This data processing pipeline makes sure that all fragments are sensible from a medicinal chemistry standpoint and carry enough information with respect to interactions to proteins. The dataset is split into training, validation and test following the approach used for HoloProt~\citep{somnath2021multi}, which is based on precomputed 30\% sequence similarity. For more detailed information we refer to Appendix~\ref{si:data}.\par 
The protein-fragment encoder is trained on individual fragments paired with the respective contacting protein surface ($\leq5$~\si{\angstrom}) and the protein ligand interaction profiler~(PLIP)~\citep{plip} assigns interactions. The generative model is trained on the same dataset but with all fragments originating from a ligand as one data point. We further make use of the data augmentation outlined above, in which fragments below the defined size threshold are recombined, leading to multiple possible fragment combinations per ligand. We further restrict the protein surface to the pocket by discarding points further than 7~\si{\angstrom} from the full ligand.

%%%%%%%%%%%%%%%%%%%%%%%%%%%%%%%%%%%%%%%%%%%%%%%%%%%%%%%
\section{Results \& Discussion}
\label{sec:res_dis}

\subsection{Latent Representation}
\begin{figure}[ht!]
\begin{center}
\includegraphics[width=\textwidth]{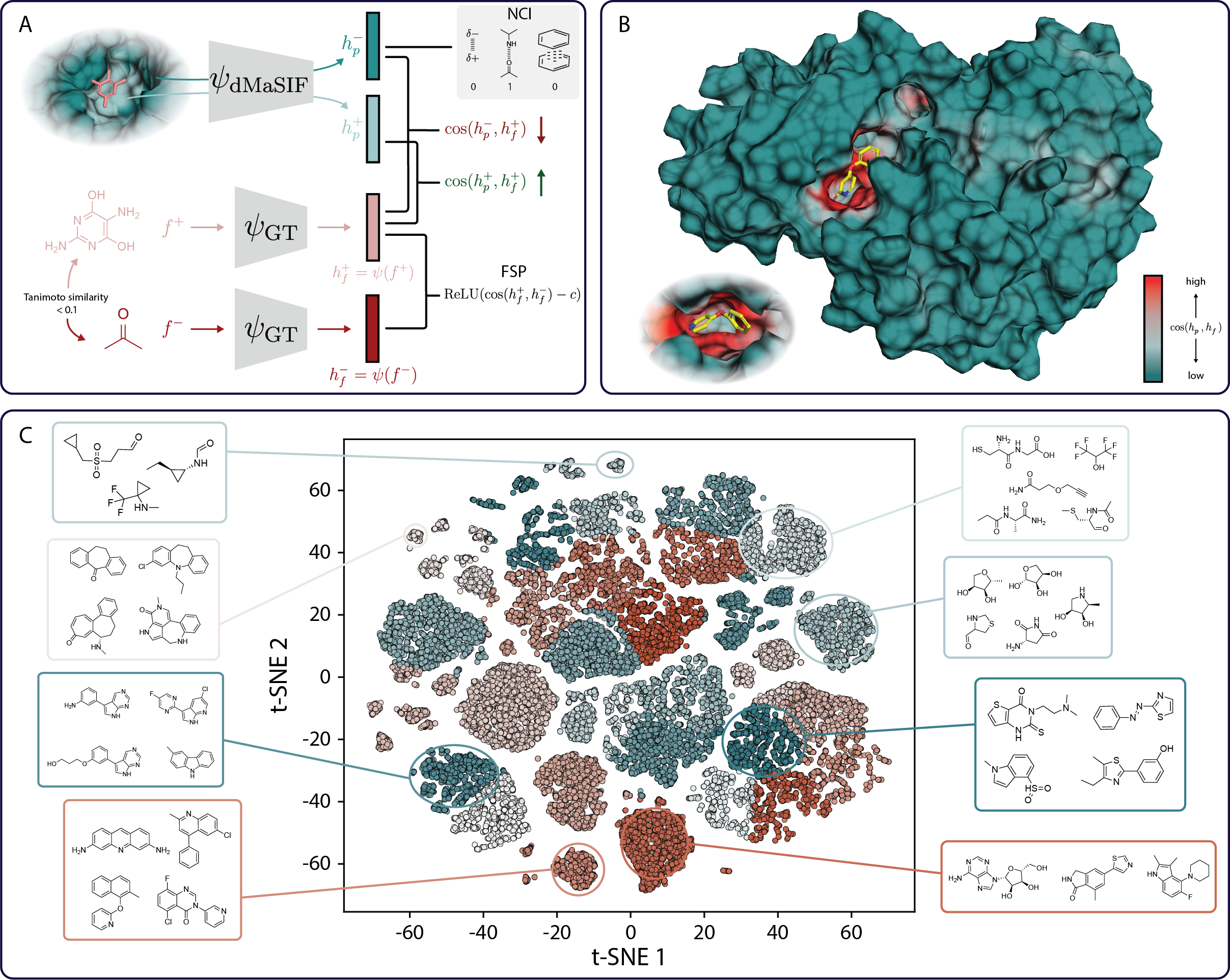}
\end{center}
\caption{\textit{A:}~Encoder pipeline, maximizing the cosine similarity $\cos(\cdot)$ between fragment $h_f$ and contacting pocket surface embeddings $h_p$ while minimizing the similarity between negative pairs. Additionally, a fragment similarity penalty~(FSP) and loss on classifying potential non-covalent interactions~(NCI) from protein surfaces are incorporated. \textit{B:}~Test set protein surface coloured by cosine similarity of every point to the fragment visually showing the sensitivity of this representation (PDB ID: 6Q4I). \textit{C:}~\mbox{t-SNE} dimensionality reduction of the fragment library in latent space with representative molecules.}
\label{fig:enc}
\end{figure}

We evaluated the quality of the learned embeddings by studying whether the similarity between the protein and fragment embeddings allows to recover the binding region of the fragment. True binding regions are defined as protein surfaces within 3~\si{\angstrom} of the fragment. To quantify this, we computed cosine similarities between a fragment and its target surface embeddings and applied several metrics: ROC AUC~(area under the receiver operating characteristic curve), AUPR (area under the precision recall curve) and $\text{EF}_1$~(enrichment factor in the 1\ts{st} percentile). We performed this study for both the whole protein surface and the pocket surface only (Table~\ref{tab:encoder_metrics}). All metrics are compared against a baseline of randomly selecting points on the protein/pocket surface as positive (binding) regions. 

The ROC AUC indicates the ability to globally discriminate between interacting and non-interacting surface regions, with a value of 0.5 corresponding to random guessing. Our encoder achieves good performances on both the whole protein~(0.76) and the pocket-level only~(0.92). The $\text{EF}_1$ measures how much more true points are identified in the 1\ts{st} percentile compared to random selection ($\text{EF}_1=1$) and thus assesses early retrieval quality compared to global ranking abilities measured by ROC AUC. The high 22.85-fold enrichment over random for the whole surface demonstrates that the model is capable of clearly distinguishing the pocket from the rest of the surface. Identifying the specific binding region within a pocket is a much harder task, which is reflected in the much lower $\text{EF}_1$ of 2.28 for the pocket. However, a 2.28-fold enrichment over random still highlights that even in the pocket itself the model chooses binding surface points twice as likely as non-interacting areas. The AUPR of 0.39 substantially exceeds the baseline of 0.16, which corresponds to the proportion of true interacting points in the pocket only. This marked improvement further highlights the model’s ability to prioritise pocket surface points over random chance. Given the imbalanced nature of binding vs. non-binding regions, AUPR is a particularly appropriate performance metric, as it demonstrates the model’s precision in identifying true positives among a large number of negatives and thus highlights the strength of our learned representation in identifying regions of potential fragment binding. 

We found that removing the FSP loss did not affect sensitivity, but led to higher average cosine similarity among all library fragments (0.45 vs. 0.25), indicating reduced diversity in the learned embeddings. This loss of embedding diversity, which we interpret as a form of mode collapse, can be detrimental for downstream fragment screening tasks. In this context, the FSP loss plays a critical role in encouraging higher entropy and chemical specificity in the fragment representations ensuring that the model does more than merely distinguish between surfaces and fragments, but actually learns chemically meaningful differences necessary for selecting appropriate fragments for novel pockets (see preliminary results in Figure~\ref{si:fig:fsp_prelim}). The effect of the NCI loss is less pronounced but does increase the metrics discussed above. This additional data can be found in Table~\ref{si:tab:enc_metrics}. 

\begin{table}[t]
\caption{Metrics evaluating the encoder on the test set and comparing to the theoretical random baseline. All scores are computed on the cosine similarity between fragment and protein surface representations with points labelled as true being 3~\si{\angstrom} away from the fragment.}
\label{tab:encoder_metrics}
\centering
\begin{tabular}{llccc}
\toprule
\textbf{Method} &\textbf{Surface set} & \textbf{ROC AUC $\uparrow$}                & \textbf{$\text{EF}_1 \uparrow$}     & \textbf{AUPR $\uparrow$} \\ \midrule
\multirow{2}{*}{encoder}&whole       &0.92 & 22.85      & 0.31   \\
&pocket      &   0.76& 2.28      & 0.39\\ \midrule
\multirow{2}{*}{random}&whole&0.5&1.00&0.01\\
&pocket&0.5&1.00&0.16\\
\bottomrule
\end{tabular}
\end{table}

We further investigated different similarity metrics on a random subset of 1000 fragments. The cosine similarity between each latent representation correlates moderately to Tanimoto similarity (Spearman $\rho=0.45$) based on the \texttt{RDKit} fingerprint (2048 bits) while there is only moderate correlation to shape and pharmacophoric similarities (after alignment) as determined by ROSHAMBO~\citep{atwi2024roshambo} (for details see Appendix~\ref{si:eval}). Molecular shape is based on overlapping atom-centred Gaussians and pharmacophores are structural features relevant for interaction with a protein such as hydrogen bond donors or aromatic rings. Interestingly, the correlation to the pharmacophore similarity is higher than the corresponding shape similarity (Spearman $\rho$ of 0.38 and 0.25 respectively) indicating that pharmacophoric aspects dominate. 

In addition to the quantitative assessment, a qualitative evaluation offers further insight into the model’s capabilities. Figure~\ref{fig:enc}B illustrates how our embeddings localise interacting surfaces with relatively high sensitivity, with the highest similarity observed between the binding fragment and its corresponding protein region. This "hotspot" can be found without any input on relative position of the two molecules or the 3D structure of the fragment. This indicates that the combination of these two latent representations captures aspects of their interaction suggesting that it can be utilised for downstream applications such as virtual screening.

We further assess the learned latent space using t-SNE dimensionality reduction demonstrating a structured and interpretable embedding space (Figure~\ref{fig:enc}C). Compared to standard \texttt{RDKit} fingerprints~(Figure~\ref{si:fig:tsne_fp}), our embeddings form distinguishable clusters, which correspond to fragments with similar chemical profiles.

\subsection{Fragment Identification}

\begin{figure}[ht!]
\begin{center}
\includegraphics[width=\textwidth]{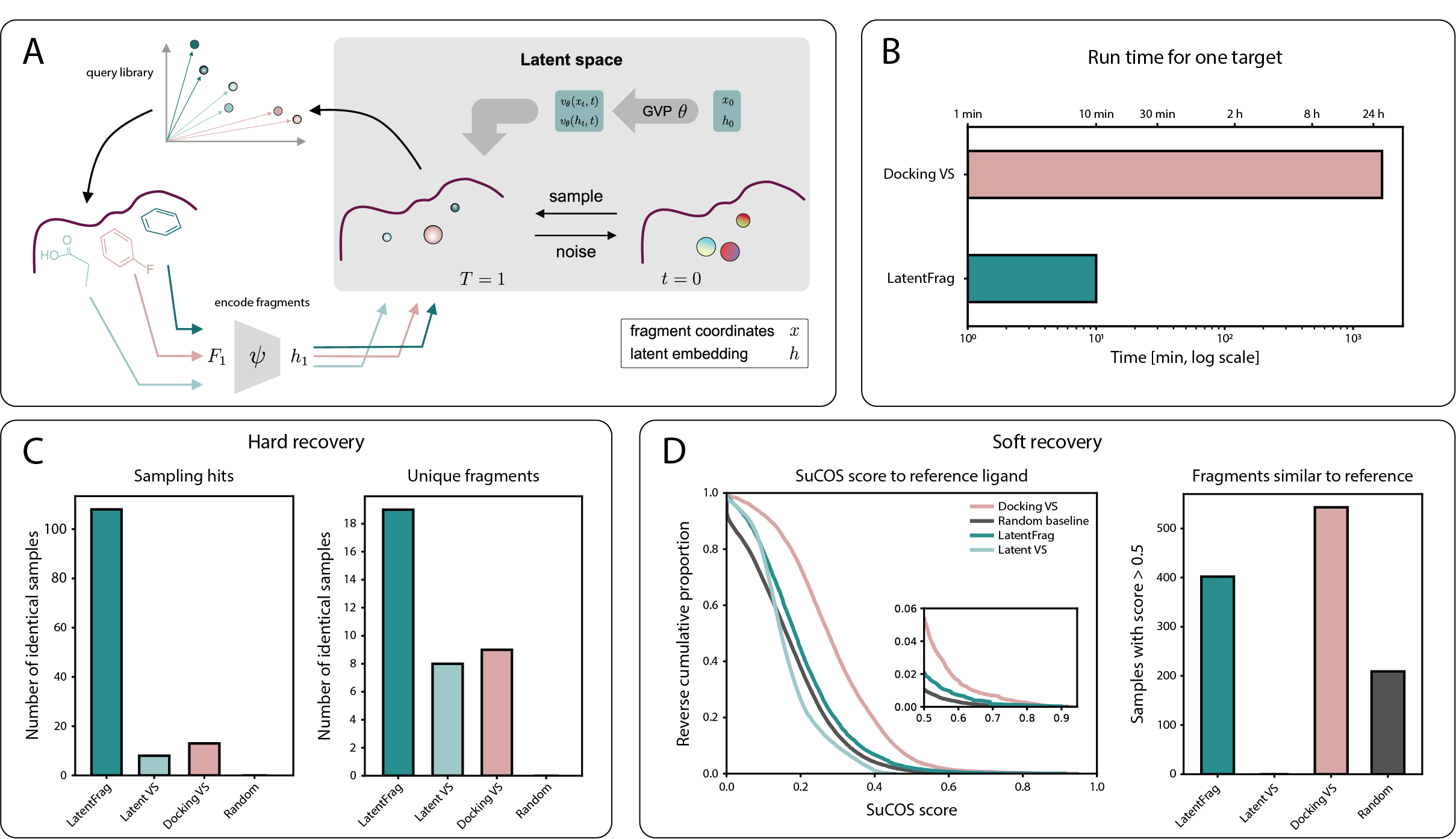}
\end{center}
\caption{\textit{A:} LatentFrag: Generative pipeline starting with encoding and the flow matching model operating in latent space. \textit{B:} Comparison of averaged rounded run times for one single target between LatentFrag and Docking VS. Bars indicate runtime in minutes in logarithmic scale (bottom axis) while the top axis shows equivalent times in linear time. \textit{C:} "Hard" recovery metrics for different sampling strategies. Both number of sampling hits (\textit{left}), which captures how many of the sampled fragments recover the ground truth and unique fragment recovery (\textit{right}), describing the number of unique recovered fragments, are shown. The RMSD refers to the distance between centroids. \textit{D:} Soft recovery metrics based on the SuCOS score relative to the full reference ligand are shown as reverse cumulative proportion (\textit{left}) and the number of fragments to the reference with a threshold of~0.5 (\textit{right}). \textit{Sampling strategies:} \textbf{LatentFrag} makes use of our generative framework, \textbf{Latent VS} screens the whole fragment library based on embedding similarity to the pocket, \textbf{Docking VS} makes use of docking scores to screen the fragment library and finally \textbf{random} samples fragments randomly.}
\label{fig:fm}
\end{figure}

A straightforward application of our learned latent space is virtual screening, focusing on its ability to identify relevant fragments by similarity to the target pocket surface. We call this sampling approach Latent VS. LatentFrag is thus a natural extension of this approach making use of generative modelling, which enables the sampling of novel fragment embeddings directly in the learned space. We compare our generative pipeline LatentFrag to Latent VS, to virtual screening based on docking scores obtained with Gnina~\citep{mcnutt2021gnina}, dubbed Docking VS, and a random baseline. For detailed information on evaluation and metrics we refer to Appendix~\ref{si:eval_ident}.

% soft and hard recovery
To evaluate the performance of LatentFrag we used hard recovery~(Figure \ref{fig:fm}C) which quantifies exact matches to reference fragments and soft recovery~(Figure \ref{fig:fm}D) which tries to assess the overlap in terms of shape and pharmacophoric patterns to the reference. We chose recovery as our main metric over frequently used docking scores because it directly reflects the model’s ability to retrieve chemically relevant fragments that are known to bind. LatentFrag consistently outperformed latent and Docking VS while random sampling failed to recover any reference fragments. This resulted in a recovery rate of LatentFrag more than double the Docking VS rate (Tab.~\ref{si:tab:recovery_rates}). Importantly, LatentFrag not only recovered more unique fragments but also did so repeatedly, achieving a sampling hit rate more than four times higher than Docking VS. These results highlight the generative model’s ability to sample relevant chemical space. It is worth noting, that the different approaches result in large differences in the number of sampled/top fragments. For Latent VS we select the top 100 fragments based on the sum of cosine similarities to the pocket and similarly for Docking VS the top 100 based on docking scores. However, for the random baseline and generative modelling we sample 100 times as many fragments as there are in the reference. This is accounted for in the calculation of the recovery rate. Notably, LatentFrag’s sampling hit rate (0.554\%) for hard recovery approaches the range of experimental fragment screening hit rates (1–2\%) at a fraction of the time and resource cost~\citep{jalencas2024design}. This suggests that while the recovery of chemically unique fragments may still be limited, the generative model is effectively enriching for relevant chemical matter and obtaining initial fragment hits to narrow the experimental search. \par 

Given the expected challenging nature of exact fragment matches, we evaluated similarity to the complete reference ligands by shape and pharmacophoric feature alignment using the SuCOS score~\citep{leung2019sucos}. With the random baseline as a reference (Figure~\ref{fig:fm}D), the generative samples demonstrate the ability to recover fragments that partially align with reference pharmacophore profiles. Interestingly, compared to the poor results on true recovery, Docking VS performs better than the other methods on the soft metric with increased number of hits. The higher number of atoms on average per fragment compared to generative sampling (19 and 13 respectively, see Fig~\ref{si:fig:natoms_hist}) as well as more realistic placement in the pocket might influence this outcome. One major advantage of our approach compared to Docking VS is speed: docking the full library to one single target takes approximately 28~h while generative sampling takes around 10~min only (Figure \ref{fig:fm}B). Moreover, once sampled, different fragment libraries or library expansions can be tested retrospectively by simply querying the new library based on the previously sampled embeddings. However, for docking, this further increases computational cost.\par

\begin{figure}[t]
\begin{center}
    \includegraphics[width=\textwidth]{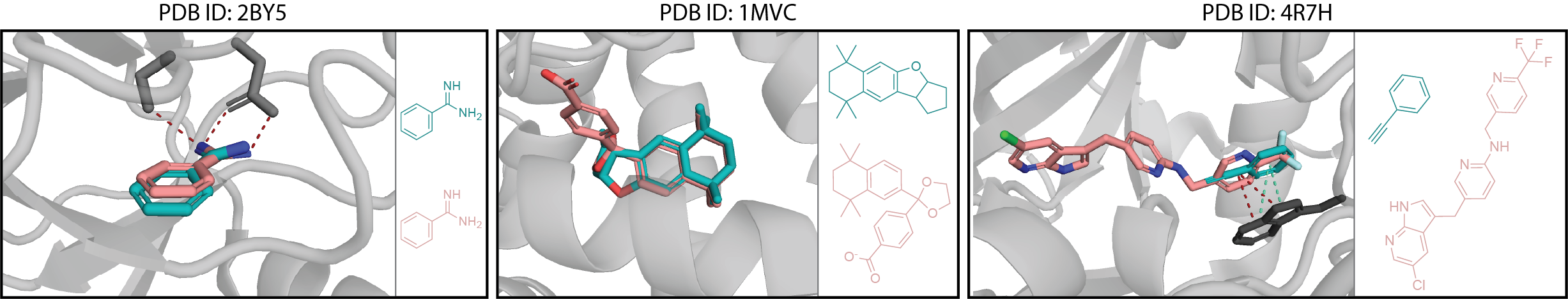}
    \caption{Selected LatentFrag samples (\textit{teal}) to test set pockets compared to the respective reference ligand (\textit{light red}) with corresponding reference interactions (\textit{red lines}). \textit{Left:} Sample that recovered the ground truth, which is also the full reference ligand. \textit{Middle:} Sample with high SuCOS score. \textit{Right:} Sample with low docking efficiency including predicted $\pi$-$\pi$-stacking similar to the reference.}
    \label{fig:sampled_ex}
    \end{center}
\end{figure}

Lastly, Figure~\ref{fig:sampled_ex} shows three selected samples to test set pockets from our generative pipeline highlighting the ability to recover sensible poses even without a constrained docking approach and match interaction profiles from the reference ligands. The first example recovers the full reference ligand and given the close overlap will likely recover also the ground truth hydrogen bonds. The second sample has a high SuCOS score to the reference and visually exhibits close overlap with the reference. When selecting by docking efficiency, we observe recovering of $\pi$-$\pi$ interactions similar to the reference ligand. For additional results we refer to Appendix~\ref{si:results}.

\subsection{Case Study: proto-oncogene c-Met}

\begin{figure}[t]
\begin{center}
    \includegraphics[width=\textwidth]{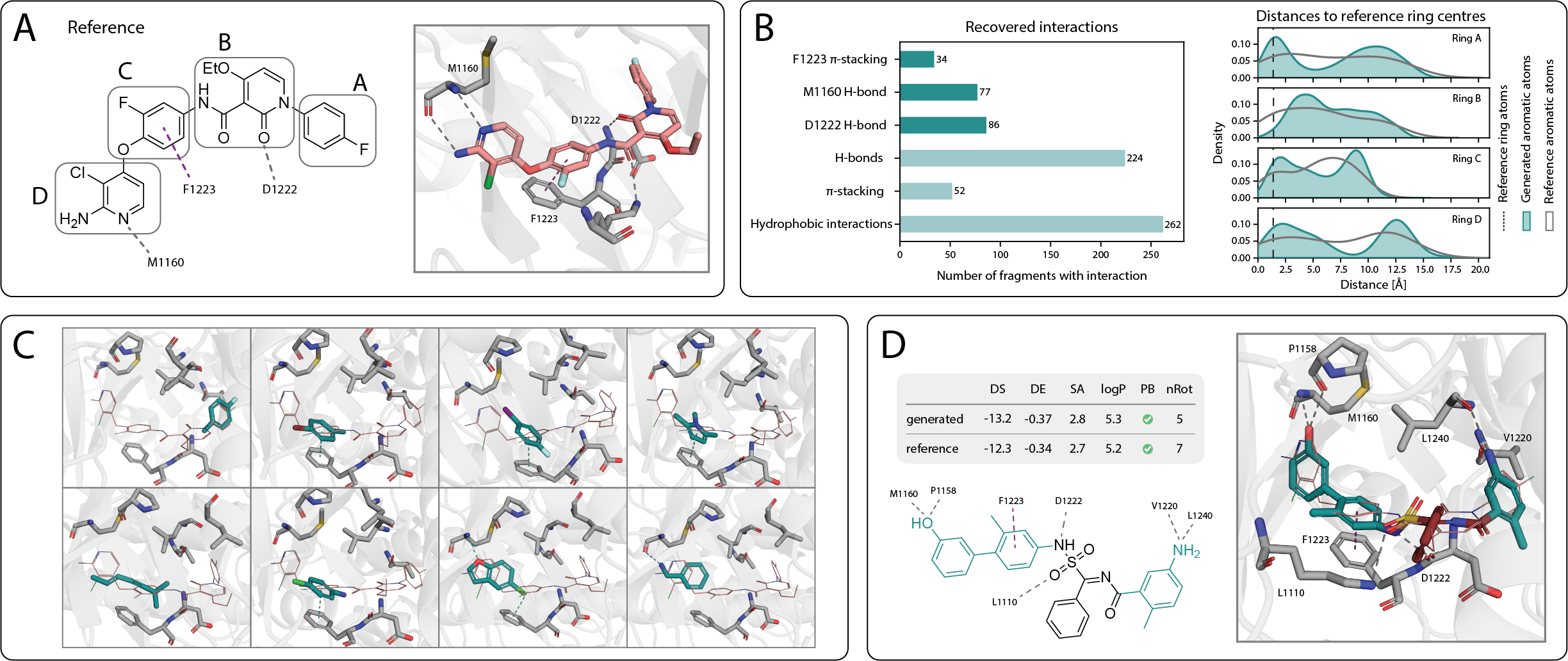}
    \caption{\textit{A}: Structure of reference BMS-777607 with hydrogen bonds (grey) and $\pi-\pi$ stacking (magenta) to its target c-Met (PDB ID:~6SDD). \textit{B}: Recovered interactions and spatial sampling of generated fragments. \textit{Left:} Number of fragments recovering the reference interactions to M1160, D1222 and F1223 as well as specific interactions (hydrogen bonds, $\pi-\pi$ stacking and hydrophobic interactions) to any residue in the pocket. \textit{Right:} Kernel density estimates of distances between generated aromatic atoms (teal) and the centres of reference rings of moieties A-D compred to the aromatic atoms of the reference (grey). The dashed black line shows the distance of the ring atoms of the respective reference rings to their centre. More detailed distributions are shown in Figure \ref{si:fig:case_study_kde}.
    \textit{C}: Selected LatentFrag sampled fragments (teal) with top (low) docking efficiencies superimposed on the reference (magenta). \textit{D}: Sampled fragments (teal) connected by a DiffLinker generated linker (black) compared to the reference ligand (magenta). The table shows the docking score (DS), docking efficiency (DE), synthetic accessibility score (SA), octanol-water partition coefficient (logP), PoseBusters benchmark success and number of rotatable bonds.}
    \label{fig:case_study}
    \end{center}
\end{figure}

We next showcase the potential of LatentFrag in a case study on the disease-relevant target c-Met (hepatocyte growth factor receptor). The proto-oncogene c-Met is known to promote the growth of several solid tumours. There are drugs approved for non-small cell lung cancer that inhibit c-Met but many of these inhibitors are rendered insufficient due to emergence of resistance. This resistance is attributed to mutations near the active site~\citep{zhang2014discovery, collie2019structural}. In this case study we investigate a D1228V mutant of c-Met, whose crystal structure bound to the investigational compound BMS-777607 (PDB ID:~6SDD) is in our test set. No similar ligand is present in the training set (highest Tanimoto similarity is 0.75; see Figure \ref{si:fig:case_study_sim_lig}). BMS-777607, as a type-II inhibitor, extends deep into the binding pocket beyond the active site and thus is assumed to be more active against mutated c-Met~\citep{zhang2014discovery}. Figure \ref{fig:case_study}A displays the structure of BMS-777607 and highlights four motifs with distinct functions often shared among type II inhibitors. These motifs were identified based on their role in interacting with the target and we will henceforth call them \emph{moieties} in order to distinguish from the \emph{fragments} obtained by BRICS-fragmentation. Moiety A occupies a deep hydrophobic pocket and is often \textit{para}-substituted. Moiety B usually has at least one amide group forming a hydrogen bond with D1222 and does not always have a ring. Moiety C is relatively conserved and forms a $\pi-\pi$ interaction with F1223. Moreover, moiety D forms a hydrogen bond between the pyridine Nitrogen and the backbone of M1160 and in doing so anchors the hinge region of c-Met~\citep{damghani2022type}.\par 

We sample new fragments 100 times using LatentFrag as described above. Every sample contains three fragments matching the number of fragments in the reference. Therefore, the reference fragments obtained by BRICS-fragmentation are not identical to the interaction-defining moieties A, B, C and D described above. The comparison of all generated fragments to the reference reveals that the model consistently recovered modes for moieties A, C and D preferentially sampling aromatic atoms at distances similar to the reference rings of the respective moieties (Figure \ref{fig:case_study}B right). This suggests that the model recovers similar types of chemistry in those modes. Solely the ring of moiety B shows less overlap as less aromatic rings are placed in the same position. However, moiety B does not primarily interact with the target through the aromatic ring and the hydrogen bond to D1222 is frequently recovered by sampled fragments (Figure \ref{fig:case_study}B left). Interactions to F1223 ($\pi-\pi$ stacking) and M1160 (hydrogen bond) are also recovered several times apart from many new interactions to residues the reference did not target. Figure~\ref{fig:case_study}C presents the top 10 fragments with respect to docking efficiency. The strong chemical and interaction profile similarity to the reference highlights the model's ability to sample relevant chemistry and also demonstrates that the generative model is not only recovering fragments hits but is also biased toward chemical features that are characteristic to true binders.\par 

Next, we extend our approach beyond fragment sampling to the full workflow of fragment-based drug design. By combining our method with established tools for fragment linking, we illustrate how it can be used to generate complete, drug-like ligands tailored to a target pocket. For this, we select two non-overlapping fragments from the aforementioned top fragments and make use of DiffLinker to generate 50 different linkers (for details see Appendix~\ref{si:sec:difflinker}). DiffLinker is a diffusion model able to generate linkers conditioned on the input fragments and the protein pocket~\citep{igashov2024equivariant}. All designed linkers with the exception of three recover the missing hydrogen bond to D1222 from the reference. The three samples not recovering the interaction form a hydrogen bond to L1110 instead. Figure~\ref{fig:case_study}D showcases one selected ligand with favourable profile. The novel ligand results in an improved docking score and docking efficiency compared to the reference while recovering all of the reference interactions and additionally forming new ones to L1110, V1220, L1240, and P1158. The molecule fulfils all PoseBusters~\citep{buttenschoen2024posebusters} criteria suggesting a physically valid pose and does not clash with the protein. The low synthetic accessibility score~\citep{ertl2009estimation} of 2.8 indicates low chemical complexity allowing synthesis of the compound. 
%A more in-depth analysis of the compound with the retrosynthesis tool AiZynthFinder~\citep{genheden2020aizynthfinder} reveals that no route to purchasable compounds could be found with this method. However, this failure stems from the novel linker and not the fragments generated by LatentFrag, which do have similar compounds in the used building block library and are thus purchasable. 
Lastly, our novel ligand contains less rotatable bonds compared to the reference, which is advantageous due to lowered conformational entropy. The excellent profile of the ligand designed from the generated fragments further highlights the potential of LatentFrag for the use in drug discovery.

\section{Conclusions}
\label{sec:conclusion}

% encoder
We introduce a novel protein-fragment encoder trained in a contrastive fashion that jointly learns rich representations of protein surfaces and molecular fragments in a shared latent space. Our approach captures key aspects of protein-fragment interactions while maintaining chemical relevance through fragment similarity-based penalties. We highlight the quality of our learned embeddings by demonstrating the recovery of binding sites of fragments on the whole protein with high sensitivity. Complementary, the encoder further enable both virtual screening directly using the latent embeddings and generative fragment identification.

% gen model
LatentFrag, our flow matching framework to perform generative fragment identification, correspondingly operates in this latent space sampling both fragment embeddings and their centroids. Importantly, our framework does so in the presence of the target pocket, which is not the case for many fragment based drug design approaches. LatentFrag demonstrated successful fragment recovery both when evaluating exact matches to the reference fragments and when assessing shape and pharmacophore similarity to the full reference ligand. We notably outperform virtual screening by docking, a popular \textit{in silico} approach for hit identification, in recovering known fragments. These results demonstrate its potential for providing initial hit hypotheses for experimental validation.

Importantly, LatentFrag is significantly faster than virtual screening using a popular docking tool on the full fragment library. Furthermore, our method offers the added advantage of flexible choice and expansion of the fragment library even after sampling. This flexibility is important, as building blocks can be interchanged with commercial sources such as Enamine~\citep{enamine-bb}, and fragments can be derived not only from crystal structures but also from generated conformers starting from SMILES. By continuing to rely on library-based fragments, synthesizability is maintained, which is a key practical consideration. This makes our approach a cost-effective alternative for early-stage screening in fragment-based drug design.

% case study
In a case study on the pharmaceutically relevant target c-Met, we recovered most modes of a reference molecule known to be essential for interacting with the target multiple times. Additionally, we successfully captured established interactions but also revealed additional plausible contacts not found in the reference. We then applied an established fragment-linking method to assemble these recovered fragments into an initial full-ligand hit that retained the reference interactions while adding new ones with potential to improve potency.

Beyond fragment identification, our surface embeddings could also be explored for binding site prediction. The results indicate that the protein surface embeddings capture features that are highly relevant for this task, providing an additional critical application.

% limitations
There remain limitations and clear directions for improvement. A key limitation is the need to know the binding pocket \textit{a priori}, which may restrict applicability in some settings. The placement of fragments through flow matching, while sufficient for fragment-based drug discovery where recovered fragments can be tested experimentally, may not yet be optimal. Furthermore, reliance on docking tools increases complexity and interpretability of the approach. By only allowing slight refinement of the placed fragment rather the full-pocket docking we aim at decreasing negative impact stemming from unspecific placement by the docking tool. Future improvements should refine the generative pipeline by implementing rigid-body transformations of fragments, improving spatial alignment, and potentially reducing reliance on docking tools. Moreover, while our evaluation demonstrates strong fragment recovery, the recovery ratio itself leaves room for improvement, and metrics based solely on recovery do not fully capture the potential of the sampled fragments. A two-stage approach involving fragment recovery followed by ligand design will likely remain necessary.

\subsubsection*{Code availability}
All the code and data can be found on \url{https://github.com/rneeser/LatentFrag}.

\subsubsection*{Acknowledgments}
R.M.N. thanks VantAI (USA) for their support and helpful feedback. I.I. has received funding from the European Union's Horizon 2020 research and innovation programme under the Marie Skłodowska-Curie grant agreement No 945363. M.B. is partially supported by the EPSRC Turing AI World-Leading Research Fellowship No. EP/X040062/1 and EPSRC AI Hub No. EP/Y028872/1. P.S. acknowledges support from the NCCR Catalysis (grant number 225147), a National Centre of Competence in Research funded by the Swiss National Science Foundation. This work was supported by the Swiss National Science Foundation grant 310030\_197724. 

\bibliography{ref}
\bibliographystyle{iclr2025_conference}

\newpage

\appendix

\setcounter{section}{0}
\setcounter{table}{0}
\setcounter{figure}{0}
\setcounter{equation}{0}
\renewcommand{\thesection}{\Alph{section}} 
\renewcommand{\thefigure}{\thesection.\arabic{figure}}
\renewcommand{\thetable}{\thesection.\arabic{table}}
\renewcommand{\theequation}{\thesection.\arabic{equation}}

\title{\textbf{Supplementary Information:} \\ Flow-Based Fragment Identification via Binding Site-Specific Latent Representations}

\maketitle

%%%%%%%%%%%%%%%%%%%%%%%%%%%%%%%%%%%%%%%%%%%%%%%%%%%%%%%
%%%%%%%%%%%%%%%%%%%%%%%%%%%%%%%%%%%%%%%%%%%%%%%%%%%%%%%

\section{Latent Encoder}
\label{si:encoder}

\subsection{Model Architecture}
\label{si:enc_architecture}

\subparagraph{Protein Encoding}
The protein encoding approach is similar to the dMaSIF~\citep{dmasif} method with some notable exceptions: a smaller receptive field $r$ (geodesic radius), higher embedding dimension $d$ and surface computation as in the original MaSIF~\citep{masif}. The protein surface is first computed using the MSMS program~\citep{msms} with a density of 3~\si{\angstrom}, a water probe radius of 1.5~\si{\angstrom} and subsequent downsampling to a resolution of 1~\si{\angstrom}. Input features computed on this surface include geometric properties (shape index, curvature) and chemical features (electrostatics, hydrogen bond donors/acceptors, hydropathy). The architecture is based on a special CNN which uses Gaussian kernels defined in a local geodesic coordinate system. We refer to~\citet{masif} for detailed information. 

\subparagraph{Fragment Encoding}
Fragments are represented as 2D graphs. Node features are initialised by one-hot encoded atom types considering \{C, N, O, S, B, Br, Cl, P, I, F, NH, N+, O-\} with NH corresponding to a Nitrogen with an explicit Hydrogen and +/- represent formal charges. Additionally, a suite of properties are computed for atoms, bonds and on the global level, which are summarised in Table~\ref{si:tab:graph_feats}. The fragment encoding is parametrised using a Graph Transformer~(GT)~\citep{dwivedi2020generalization, vignac2022digress, retrobridge} directly using the output of the learned global embedding $y$. We largely follow the GT implementation of~\citet{retrobridge} with one notable exception of linearly projecting the input features before adding them to the network output instead of simply cropping the vector.
% XXX if have time: write out math in more detail

\begin{table}[H]
  \caption{Features for initializing nodes, edges and global features of the fragment graph. Non-binary atom and bond properties were one-hot encoded and global continuous values min-max scaled. These properties were determined using \texttt{RDKit}.~\citep{rdkit}}
  \label{si:tab:graph_feats}
  \centering
  \begin{tabular}{llll}
    \toprule
    \textbf{Localization}     & \textbf{Feature}     & \textbf{Values} & \textbf{Size} \\
    \midrule
    \multirow{7}{*}{atom}& atom type  &  \{C, N, O, S, B, Br, Cl, P, I, F, NH, N+, O-\} & 13\\
    & formal charge & [-1,3] & 5 \\
    & degree & [0,6] & 7 \\
    & is in ring & binary & 1 \\
    & is aromatic & binary & 1 \\
    & hybridization & \{sp, sp2, sp3, sp2d, sp3d, sp3d2, unspecified\} & 7 \\
    & chiral tag & [S, R, unspecified, other] & 4\\ \midrule
     & bond type & \{single, double, triple, aromatic, no bond\} &5\\
     bond & conjugated & binary & 1\\
     & is in ring & binary & 1 \\ \midrule
     \multirow{10}{*}{global} & \# atoms & continuous& 1\\
     & \# bonds &continuous & 1\\
     & \# rings  & continuous& 1\\
     & \# aromatic rings &continuous & 1\\
     & MW & continuous& 1\\
     & logP &continuous & 1\\
     & TPSA & continuous& 1\\
     & \# HBD &continuous& 1 \\ 
     & \# HBA &continuous& 1 \\
     & ring sizes &[3, 18], no ring, other ring size & 18\\\bottomrule
  \end{tabular}
\end{table}

\subsection{Contrastive Training}
The encoder is trained contrastively maximizing the cosine similarity for positive protein-fragment pairs while minimizing it for negative pairs sampled from the same protein pocket.

\subparagraph{Positive and Negative Pairs}
Positive surface points $\mathcal{P}^+$ are defined as all surface points within a threshold distance $d_{bind}$ of any fragment atom. Negative surface points $\mathcal{P}^-$ are sampled from various regions within the pocket, excluding positive points. These include:
\begin{itemize}
    \item Random pocket points: Points sampled uniformly from the pocket.
    \item Concave points: Predicted by a curvature classifier as regions with inward curvature.
    \item Convex points: Predicted as regions with outward curvature.
\end{itemize}

To classify surface curvature as concave or convex, we train a Histogram-based Gradient Boosting Classifier from \texttt{scikit-learn}~\citep{curv_classifier} on the same dataset using curvatures at different smoothing scales as described in dMaSIF~\citep{dmasif}. The classifier labels concave regions based on their proximity to fragments ($d<d_{bind}$), assuming such regions are more likely to host binding interactions, while convex points are those outside this threshold. The classifier is trained and run concurrently with the encoder training, enabling on-the-fly curvature prediction.

\subparagraph{Contrastive Loss}
For the contrastive training loss, the cosine similarity between the protein surface embedding $h_p$ and fragment embedding $h_f$ is computed as follows:
\begin{equation}
    \cos(h_p,h_f)=\frac{h_p\cdot h_f}{\|h_p\|\|h_f\|}.
\end{equation}
Positive pairs maximise $\cos(h_p,h_f)$ while negative minimise it
\begin{align}
    \mathcal{L}_{pos}&=-\mathbb{E}_{(p,f)\in\mathcal{P}^+}\left[w(p,f)\log\sigma(\cos(h_p,h_f))\right] \\
    \mathcal{L}_{neg}&=-\mathbb{E}_{(p,f)\in\mathcal{P}^-}\left[\log\sigma(-\cos(h_p,h_f))\right].
\end{align}
Here, $\sigma$ is the sigmoid function, and $w(p,f)$ scales the contribution of positive pairs by their distances $d$, defined as:
\begin{equation}
    w(p,f)=\frac{1}{1+\exp(-\alpha(\frac{d-d_{min}}{d_{max}}-0.5))}
    \label{si:eq:weight}
\end{equation}
where $d_{min}$ and $d_{max}$ normalise the distances, and $\alpha$ controls the sharpness of the weighting. This up-weights positive points that are further away, encouraging the model to focus on points at the edge of the binding regions. This was shown to work much better than the more intuitive way of penalizing far points by inverting the sign before $\alpha$ in Equation~\ref{si:eq:weight}, which surprisingly resulted in approximately random ROC AUC scores on the test set. This weighting scheme also slightly improves over the binary case (no weights) especially with respect to discriminative metrics such as $\text{EF}_1$ or top-$k$ Accuracy also on the test set. One potential explanation for this is that interactions further away are weaker albeit more numerous, which in return can result in them being "drowned out" and the model not taking this collective effect of interactions into account.\par 
The final contrastive loss is the sum of both terms

\begin{equation}
    \mathcal{L}_{contrastive}=\frac{\mathcal{L}_{pos}+\mathcal{L}_{neg}}{2}
\end{equation}

\subsection{Fragment Similarity Penalty (FSP)}
To incorporate aspects purely dependent on the small molecular fragment and encourage diversity of embeddings we introduce a fragment similarity penalty~(FSP). The FSP is a hinge loss between the true positive fragment embedding $h_f^+$ and the embedding of a randomly sampled fragment from the same training library with Tanimoto similarity below 0.1. This penalises similar embeddings despite dissimilar chemical structures. The loss is defined like so
\begin{equation}
    \mathcal{L}_{FSP}=\text{ReLU}(\cos(h_f^+,h_f^-)-c)
\end{equation}
with margin $c$.

\subsection{Non-covalent Interaction (NCI) Loss}
The additional NCI loss is based on a multi-label classifier predicting non-covalent interaction (NCI) types for protein pocket surface points using their latent embeddings. This encourages the differentiation of the pocket surface and aims at increasing sensitivity to different fragments. The NCI classifier module is a lightweight feedforward neural network with one linear layers, surface embeddings as input and $n_{NCI}+1$ classes as output using a sigmoid activation function. The model is thus tasked with predicting probabilities for the classes \{hydrophobic interactions, hydrogen bonds, water bridges, salt bridges, $\pi$-stacks, $\pi$-cation interactions, halogen bonds, interaction presence\}. Interactions are extracted by using the Protein-Ligand Interaction Profiler~(PLIP)~\citep{plip} and positive labels, as one-hot encoded interaction profile, are assigned when surface points are in a distance of 2~\si{\angstrom} of the interacting residue atom. One surface point can have several interactions if it is close to multiple interactions. The loss is the the binary cross entropy~(BCE) between predicted probabilities $z$ and true labels $y$
\begin{equation}
    \mathcal{L}_{NCI}=\text{BCE}(z, y) 
\end{equation}

\subsection{Training Loss}
The final training loss is the weighted sum of the previous individual losses including L2 regularization
\begin{equation}
    \mathcal{L}=\lambda_{contrastive}\mathcal{L}_{contrastive}+\lambda_{reg}\mathcal{L}_{reg}+\lambda_{FSP}\mathcal{L}_{FSP}+\lambda_{NCI}\mathcal{L}_{NCI}.
\end{equation}
with the regularization term amounting to
\begin{equation}
    \mathcal{L}_{reg}=\frac{1}{n+m}\left(\sum_{i=1}^{n}h_{p,i}^2+\sum_{j=1}^{m}h_{f,j}^2\right).
\end{equation}

\subsection{Training}

\begin{table}[H]

% XXX parameters size
\label{si:tab:hparams_encoder}
\caption{Hyperparameters of the encoder training.}
\begin{center}
\begin{tabular}{clc}
\toprule
\textbf{Module}   & \textbf{Parameter}          & \textbf{Value}                                                                                  \\ \midrule
\multirow{3}{*}{training}                                                                        & epochs                      & 25                                                                                              \\
                                                                                & batch size                  & 32                                                                                              \\ 
                                                                                & learning rate               & 0.001                                                                                           \\ \midrule
 \multirow{10}{*}{loss}                                                          & $\lambda_{contrastive}$             & 2.0 \\
                                                                                & regularization              & 0.1                                                                                             \\ 
                                                                                & NCI loss                    & yes                                                                                             \\ 
                                                                                & $d_{min}$ [\si{\angstrom}] & 0.5\\
                                                                                & $d_{max}$ [\si{\angstrom}] &$d_{bind}=3.0$ \\
                                                                                & $\alpha$ &5.5 \\
                                                                                & $\lambda_{NCI}$  & 1.0                                                                                             \\
                                                                                & FSP loss                         & yes                                                                                             \\
                                                                                & $\lambda_{FSP}$  & 1.0                                                                                             \\ 
                                                                                & FSP margin $c$                  & 0.3                                                                                             \\ 
                                                                                & threshold $d_{bind}$ [\si{\angstrom}] & 3.0                                                                                        
                                                                                           \\ \midrule
\multirow{3}{*}{negative samples}                                               & random [\%]         & 0.34 \\  
                                                                                &concave [\%]& 0.33\\
                                                                                &convex [\%]& 0.33\\ \midrule
\multirow{6}{*}{\begin{tabular}[c]{@{}c@{}}protein encoder\\ dMaSIF\end{tabular}}                & radius $r$ [\si{\angstrom}]              & 3.0                                                                                             \\ 
                                                                                & resolution [\si{\angstrom}]          & 1.0                                                                                             \\ 
                                                                                & layers                      & 2                                                                                               \\ 
                                                                                & hidden dimension            & 16                                                                                              \\ 
                                                                                & embedding dimension         & 128                                                                                             \\ 
                                                                                & curvature scales            & \{1.0, 3.0, 5.0, 7.0 9.0\}                                                                      \\ \midrule
\multirow{9}{*}{\begin{tabular}[c]{@{}c@{}}fragment encoder\\ GT\end{tabular}} & layers                      & 4                                                                                               \\
                                                                                & input node dimension        & 37                                                                                              \\
                                                                                & input edge dimension        & 5                                                                                               \\ 
                                                                                & input global dimension      & 27                                                                                              \\ 
                                                                                & encoder dimensions          & $X$: 256, $E$: 64, $y$: 256                                                                           \\ 
                                                                                & hidden dimensions           & \begin{tabular}[c]{@{}c@{}}$dx$: 256, $de$: 32, $dy$: 256\\ $d_{ffX}$: 256, $d_{ffE}$: 64, $d_{ffy}$: 256\end{tabular} \\ 
                                                                                & heads                       & 8                                                                                               \\ 
                                                                                & output node dimension       & 128                                                                                             \\ 
                                                                                & output global dimension     & 128                                                                                             \\ \bottomrule                                                                                            
\end{tabular}
\end{center}
\end{table}

%%%%%%%%%%%%%%%%%%%%%%%%%%%%%%%%%%%%%%%%%%%%%%%%%%%%%%%

\section{Generative modelling}
\label{si:fm}
The architecture and flow matching framework was inspired by DrugFlow~\citep{schneuing2024towards} and \citet{schneuing2025multidomain}.

\subsection{Flow matching} 
Flow matching is a generative method, where a prior $p_0$ is connected to the data distribution $p_1$ through a sequence of probability distributions $\{p_t:t\in [0,1]\}$. This path is defined by a time-dependent vector field that can be approximated by this method. The transform the prior into a data point the vector field $u_t(x)$ is integrated to give the flow
\begin{equation}
\label{eq:ode}
    \frac{d}{dt}\psi_t(x)=u_t(\psi_t(x)).
\end{equation}
\citet{lipman2022flow} showed that rather than defining the true vector field $u_t(x)$, it is easier to parametrise the conditional flow $u_t(x|x_1)$ based on a data point $x_1$. The conditional flow matching loss is defined as
\begin{equation}
    \mathcal{L}_{CFM}=\mathbb{E}_{t,q(x_1),p_t(x_0)}\|v_\theta(t,x_t)-\dot{x}_t\|^2.
\end{equation}
with $\dot{x}_t|=\frac{d}{dt}\psi(x_0|x_1)$, the time derivative of the conditional flow.
To generate new samples, we sample $x_0$ from the prior distribution $p(x_0)$ and simulate the ODE in equation~\ref{eq:ode} using the learned vector field $v_\theta(t,x_t)$.

\subsection{Centroid coordinates: Euclidean Flow} 
To parametrise the centroid coordinates we use the Independent-coupling conditional flow matching~\citep{albergo2022building, tong2023improving} with the generative vector field
\begin{equation}
    u_t(x|x_1)=\frac{\sigma'_t(x_1)}{\sigma_t(x_1)}(x-\mu_t(x_1))+\mu'_t(x_1).
\end{equation}
A constant velocity vector field $\dot{x}_t=\frac{x_1-x_t}{1-t}$ results from a constant $\sigma=\sigma_t(x_1)$ and $\mu_t(x_1)=tx_1+(1-t)x_0$ to define the Gaussian probability path. The loss for this flow thus amounts to
\begin{equation}
    \mathcal{L}_{coord}=\mathbb{E}_{t,q(x_1),p_t(x_0)}\|v_\theta(t,x_t)-(x_1-x_0)\|^2.
\end{equation}

\subsection{Latent Fragment Embeddings: Spherical flow} 
For the latent embeddings of the fragments, we define a flow on the unit sphere $\mathbb{S}^2=\{x\in\mathbb{R}^{2+1}:\|x\|_2=1\}$. We can interpolate between $x_0$ and $x_1$ in the tangent space in which the vector field $v_\theta(t,x_t)$ is also learned, allowing one to avoid simulation. Concurrently, the loss is also computed in tangent space where the local geometry is Euclidean
\begin{equation}
    \mathcal{L}_{S^2}= \mathbb{E}_{t,q(x_1),p_t(x_0)}\|v_\theta(t,x_t)-\frac{\log_{x_t}(x_1)}{1-t}\|^2
\end{equation}

To obtain $x_t$ as the point on the sphere, we employ exponential and logarithmic maps

\begin{equation}
\label{eq:spherical_xt}
    x_t=exp_{x_0}(t\text{log}_{x_0}(x_1))
\end{equation}
with $\exp_x(u)=\cos(\|u\|)x+\sin(\|u\|)(\frac{u}{\|u\|})$, which maps a tangent vector $u$ to the sphere and $\log_x(y)=\frac{\theta}{\sin(\theta)}(x-\cos(\theta)y)$, which takes point $x$ on the sphere to a vector in the tangent space at base $y$. $\theta=\arccos(<x,y>)$, the geodesic distance between $x$ and the base $y$. Equation \ref{eq:spherical_xt} corresponds to the spherical linear interpolation~(SLERP)~\citep{slerp}
\begin{equation}
    x_t=\text{SLERP}(x_0,x_1,t)=\frac{\sin((1-t)\theta)}{\sin(\theta)}x_0+\frac{\sin(t\theta)}{\sin(\theta)}x_1.
\end{equation}

The spherical flow allows consistency in relation to the cosine similarity-based encoder training and library searching and results in a smooth trajectory between $x_0$ and $x_1$. A Euclidean flow, while also applicable, exhibits a drastic change in cosine similarity only with high $t$ requiring the model to learn big velocities for later time points only. This flow is similar to the system described by~\citet{spherical_theory}.

\subsection{Training Loss}
Combining the two modalities gives a weighted sum of the previously defined losses:
\begin{equation}
    \mathcal{L}=\lambda_{coord}\mathcal{L}_{coord}+\lambda_{S^2}\mathcal{L}_{S^2}
\end{equation}

\newpage
\subsection{Backbone Architecture}

\begin{figure}[ht!]
    \begin{center}
    \includegraphics[width=\textwidth]{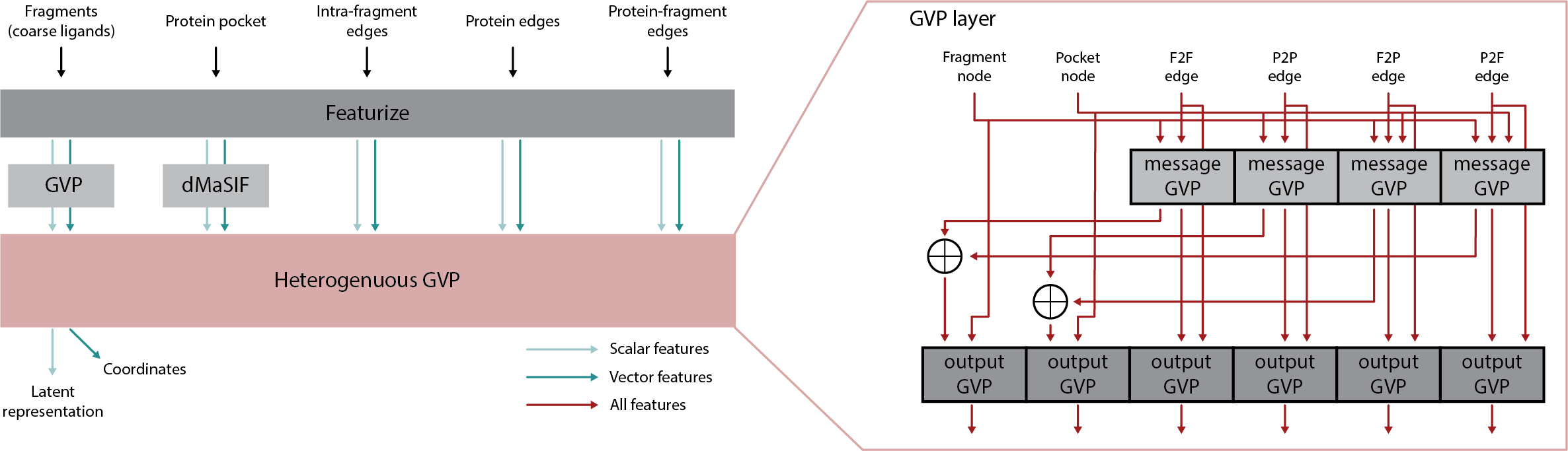}
    \caption{Schematic overview of backbone architecture of the neural network parametrizing out flows. \textit{Left:} All inputs are featurized individually to scalar and vector features, subsequently the fragment and pocket node features are transformed with a GVP and the pre-trained dMaSIF module from the encoder respectively before being passed to a shared heterogeneous GVP~\cite{jing2021equivariant}. Out output consists of velocities for the latent representation (scalar) and fragment centroid coordinates (vector). \textit{Right:} Detailed layer of the GVP with distinct messages computed based on source and destination nodes for every edge feature. All modules are processed separately in output module based on another GVP. Figure adapted from \citep{schneuing2025multidomain}.}
    \label{si:fig:GVP_scheme}
    \end{center}
\end{figure}

\subparagraph{Protein Representation}

Proteins are represented as pocket surface point clouds, restricted to a 7~\si{\angstrom} radius around the ligand. Surface points are featurized with normals as vector features and latent embeddings from the MaSIF module of the encoder as scalar features. For edge construction in the heterograph, k-nearest neighbours are sampled to connect nodes. Edge distances transformed via a radial basis function~(RBF) with 16~bases.

\subparagraph{Fragmented Ligand Representation}
Small molecule ligands are represented as coarse graphs, where fragmented ligands are decomposed into nodes and edges. Each node corresponds to a fragment, with its centroid as the position~$x$ and a latent embedding serving as the scalar node feature~$h$. No additional vector features are used except from self-conditioning. The graph is fully connected, as the goal is to learn fragment placement and identification rather than reconstruct to the ligand. Edge features are derived from pairwise distances using a RBF with 16 bases.

\subparagraph{Neural Network}
Combining pocket and fragments results in a heterogenous graph that is composed of two distinct node groups, fragments and pocket surface points, connected by four types of edges: fragment-to-fragment~(F2F), fragment-to-pocket~(F2P), pocket-to-fragment~(P2F), and pocket-to-pocket~(P2P). To accommodate this complexity, we employ a heterogeneous graph neural network architecture using geometric vector perceptron~(GVP) layers~\cite{jing2021equivariant}. This architecture incorporates distinct learnable message functions for each edge type and separate update functions for each node type. The pocket nodes are not parametrised by a GVP but instead by the MaSIF module pre-trained during the contrastive learning (weights are not frozen). Figure~\ref{si:fig:GVP_scheme} displays a scheme of the network architecture.

\subsection{Training}

\begin{table}[H]
\label{si:tab:hparams_fm}
\caption{Hyperparameters for the training of the flow matching model. s = scalar feature, V = vector feature}
\begin{center}
\begin{tabular}{clc}
\toprule
\textbf{Module}                               & \textbf{Parameter}                & \textbf{Value}   \\ \midrule
\multirow{2}{*}{training}            & epochs                   & 44\tablefootnote{Training stopped early due to convergence and limiting training time.}     \\
                                     & learning rate                       & 5e-4    \\ \midrule
\multirow{2}{*}{loss}                & $\lambda_x$               & 1.0       \\
                                     & $\lambda_h$               & 100.0     \\ \midrule
\multirow{2}{*}{heterogeneous graph} & pocket edge k neighbours & 10      \\
                                     & edge cutoff interaction [\si{\angstrom}]  & 10.0    \\ \midrule
\multirow{3}{*}{GVP}                 & layers                   & 5       \\
                                     & node hidden dimensions (s, V) & 265, 32 \\
                                     & edge hidden dimensions (s, V)   & 64, 16  \\ \midrule
simulation parameters                          & steps                    & 500  \\
\bottomrule
\end{tabular}
\end{center}
\end{table}

\subsection{Sampling and Post-processing}
\label{si:sampling}
\subparagraph{Self-conditioning}
Self-conditioning~\citep{sc} is used, in which the model takes previous predictions as input during sampling. This has been shown to improve performance. 

\subparagraph{Number of Coarse Nodes}
In order to sample new fragments, the model needs to know how many nodes to place to build a graph. This can be done either by ground truth size, given this information is available, or alternatively is made dependent on the training distribution of number of coarse nodes given a pocket with $x$ number of residues.

\subparagraph{Library Querying}
% library querying
At the end of sampling, the model outputs fragment embeddings $h$ and centroid coordinates $x$. We can move from this coarse representation to an all-atom output by querying a fragment library maximum cosine similarity of the respective latent embeddings. This fragment library is independent of sampling or training and thus flexibly interchangeable.

\subparagraph{Docking}
% XXX (maybe goes into metrics and evaluation?)
In order to obtain a realistic orientation with respect to the protein, all fragments are docked individually. Docking is performed using Gnina~\citep{mcnutt2021gnina} with the fragment placed at the sampled centroid defining the bounding box plus a margin of~1~\si{\angstrom}.

%%%%%%%%%%%%%%%%%%%%%%%%%%%%%%%%%%%%%%%%%%%%%%%%%%%%%%%
\section{Data}
\label{si:data}
% PDB extraction
For both the encoder and the generative modelling protein-ligand structures were extracted from the Protein Data Bank~(PDB)~\citep{berman2000protein}. For this, PDB was queried (accessed May 2023) retrieving 122,012 protein-only entries where the structures were determined using one of the following experimental methods: X-ray diffraction, electron microscopy, solid-state NMR, or solution NMR. The entries are filtered to include only those with a refinement resolution of 3~\si{\angstrom} or better and at least one distinct non-polymer entity present. Subsequently, all structures are split into individual chains and corresponding ligands. Ligands determined to be irrelevant to the task are removed following a list of identifiers used by Binding MOAD~\citep{hu2005binding}, which includes solvents, buffers, other crystallization artifacts etc. All ligand SMILES present in the PDB are extracted to reassign the correct bond order using \texttt{RDKit}. Pairs where this reassignment failed were discarded. All proteins were protonated using Reduced~\citep{word1999asparagine}.\par
All ligands were fragmented using BRICS rules~\citep{degen2008art} with the exception of breaking double bonds. Fragments smaller than 8 heavy atoms got recombined with all possible neighbouring fragments (related to graph partitioning), which served as a data augmentation technique.

\subsection{Encoder}
\label{si:data:encoder}

For the encoder, only fragments in the distance of at least 5~\si{\angstrom} were kept and the following filtering criteria were applied. Fragments are discarded if:
\begin{itemize}
    \item Have no interaction as determined by PLIP~\citep{plip}.
    \item Have more than 20 heavy atoms.
    \item An element present is not in \{C, N, O, S, B, Br, Cl, P, I, F\}
    \item Molecular weight is more than 500~Da.
    \item Have a maximal ring size of more than 8.
    \item Have a phosphate group or an aklyl chain of at least 4 Carbons.
\end{itemize}
The dataset is split into training, validation and test following the approach used for HoloProt~\citep{somnath2021multi}, which is based on precomputed 30\% sequence similarity. This splitting approach does result in having overlapping fragments but no data leakage in terms of protein sequence similarity. \par 
Each fragment in the dataset is randomly paired with another fragment with a Tanimoto similarity below 0.1 for the FSP.\par

\begin{figure}[ht!]
\begin{center}
\includegraphics[width=0.6\textwidth]{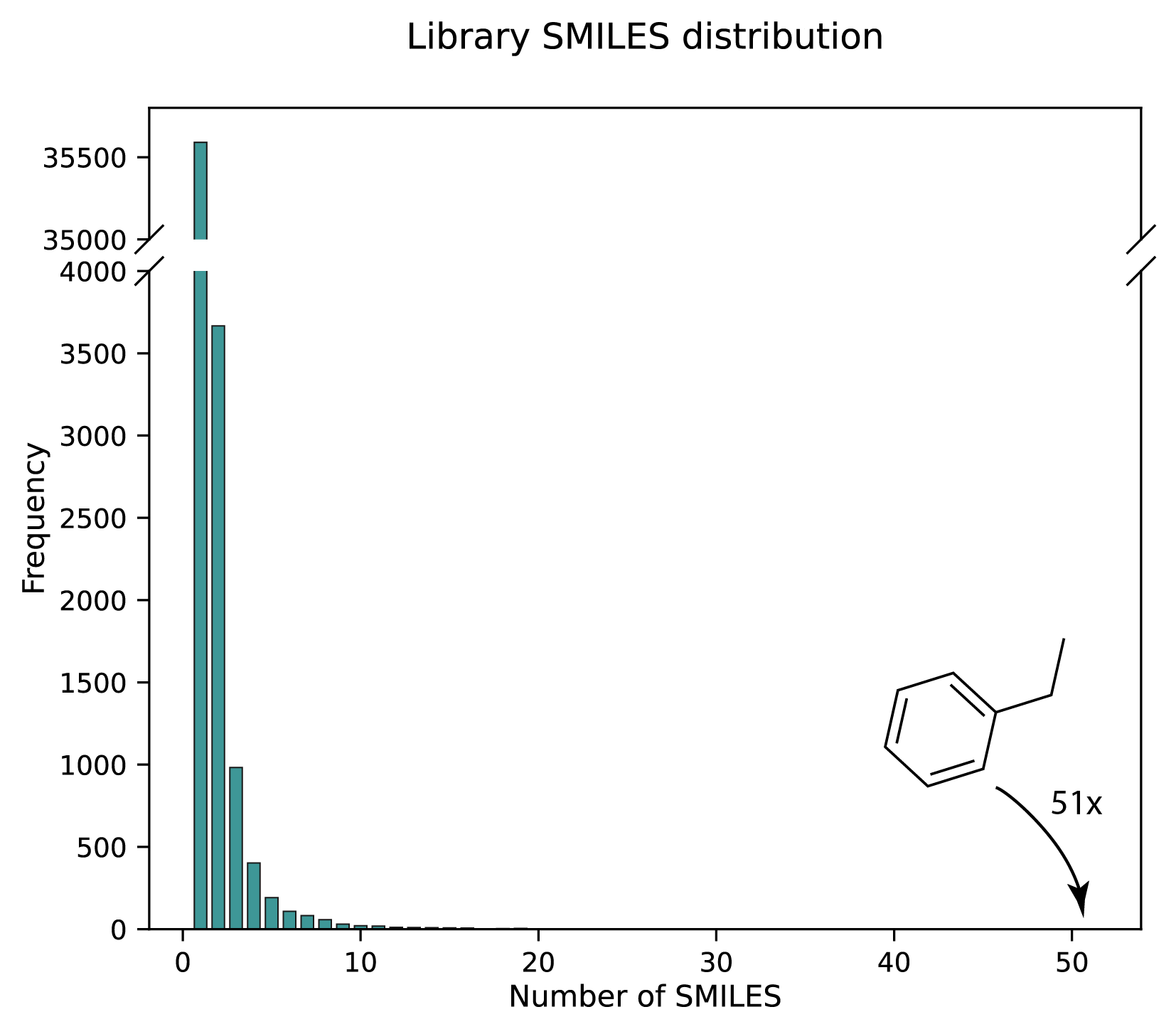}
\end{center}
\caption{Distribution of frequency of SMILES in the fragment library. Individual fragments can appear multiple times with different conformations but will have the same latent encoding. Ethylbenzene is the most frequent fragment and appears 51 times..}
\label{si:fig:smi_dist}
\end{figure}

\begin{figure}[ht!]
\begin{center}
\includegraphics[width=\textwidth]{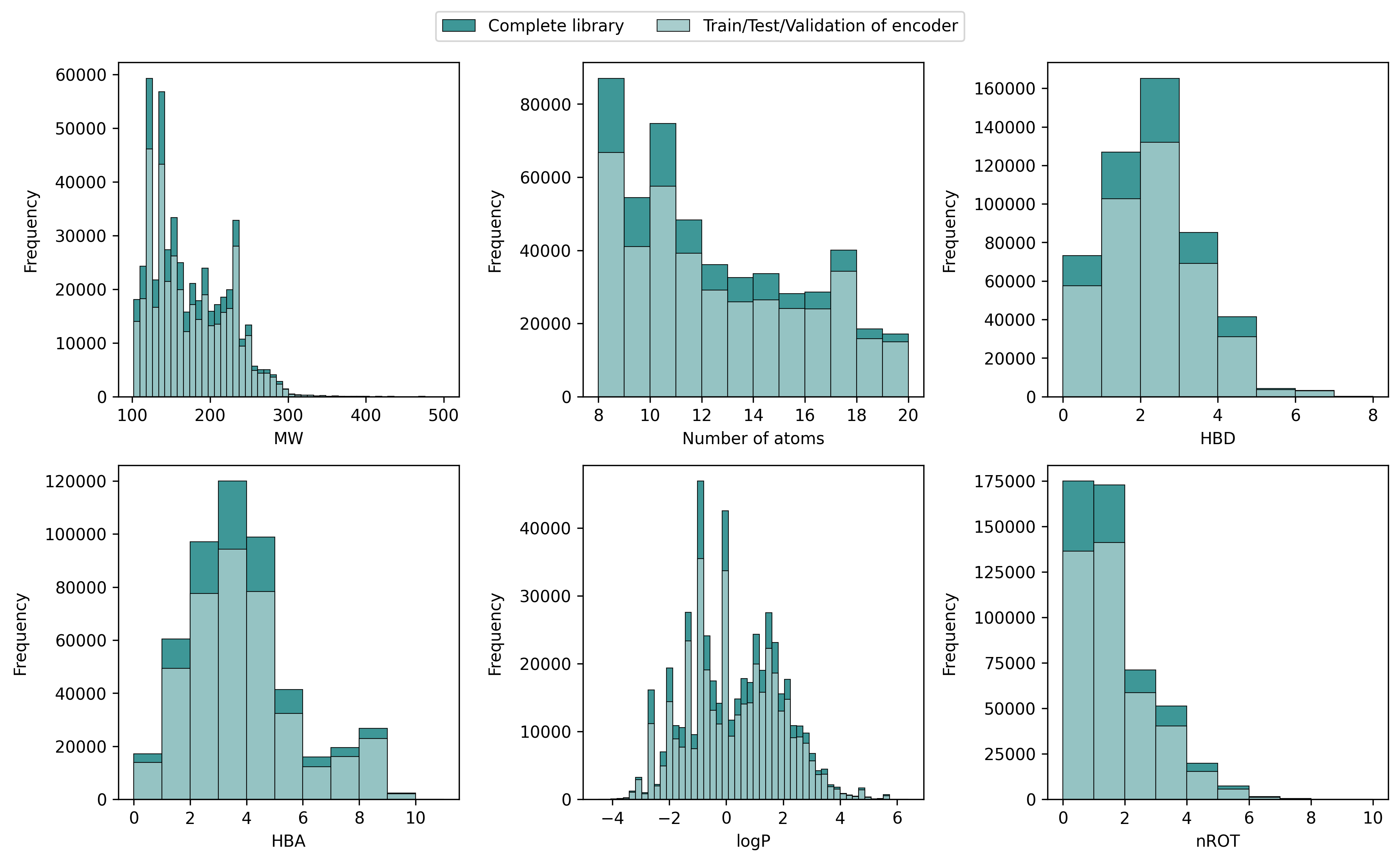}
\end{center}
\caption{Molecular properties of fragments extracted from the PDB structures. Both the complete library (\textit{dark teal}) used for querying fragments in the generative pipeline and the set filtered by the presence of interactions (\textit{light teal}) are shown. MW = molecular weight, HBD = hydrogen bond donor, HBA = hydrogen bond acceptor, logP = lipophilicity, nROT = number of rotational bonds}
\label{si:fig:lib_stats}
\end{figure}

The fragment library used for VS and the generative pipeline is identical to the training fragments with the exception of removing fragments that have no profiled non-covalent interactions. This results in a fragment library of 41,224 unique fragments and 52,070 conformers (Figure~\ref{si:fig:smi_dist}), with the lowest energy conformer being retained for downstream tasks. The full library consist of 86,927 unique chains. The training set consists of 310,298, the validation set of 33,547 and the test set 45,210 protein:fragment pairs. Figure~\ref{si:fig:lib_stats} shows the distribution of some important molecular properties for the training data and the library.

\subsection{Flow Matching}
\label{si:data:fm}
For the training and evaluation of the generative framework we use the same splits based on 30\% sequence similarity as described above but randomly subsample the validation and test set to 100 datapoints with unique pockets. We further use the same data augmentation technique as described for the encoder for which we recombine fragments smaller than 8 heavy atoms in a combinatorial manner with their neighbours in order to get big enough fragments. One datapoint will have all fragments necessary to make up one ligand and contrary to the encoder, fragments further away are not discarded. The training set comprises of 118,786 pocket:fragment-set pairs.\par 
Pocket surfaces are extracted by removing points further than 7~\si{\angstrom} from the full ligand. Pocket surfaces points cloud smaller than 250 points are discarded.

%%%%%%%%%%%%%%%%%%%%%%%%%%%%%%%%%%%%%%%%%%%%%%%%%%%%%%%
\section{Metrics and Evaluation}
\label{si:eval}

\subsection{Encoder}
\label{si:sec:metrics_encoder}

The encoder is evaluated on the full test set by assessing results obtained from both full surfaces and pocket surfaces only. The \textbf{ROC AUC} (area under the receiver operating characteristic curve) represents the main training objective of being able to discriminate between positive (interacting) and negative (non-interacting) surface:fragment pairs and is computed based on their cosine similarities across the whole surfaces.
The following metrics are all computed on the cosine similarities between a fragment and each protein surface point and true (interacting) points are defined as those within 3~\si{\angstrom} of the fragment.\par
The \textbf{enrichment factor (EF)} measures how well high-similarity points are enriched for binding sites compared to random expectation. All points above a similarity threshold given by the specified percentile (we consider top 1\textsuperscript{st}, 5\textsuperscript{th}, and 10\textsuperscript{th} percentile) are considered high-similarity points. The EF is the division of the fraction of interacting points among high similarity points and the fraction of interacting points across all points (whole surface or pocket). The \textbf{top-$k$ accuracy} evaluates how well the highest-scoring points correspond to actual binding sites. It first ranks all points based on similarity and selects the top $k$ (1, 10, 100) highest-scoring ones to compute the accuracy. The \textbf{Area Under the Precision-Recall Curve (AUPR)} quantifies how well similarity values distinguish binding from non-binding sites. The precision-recall curve is computed by varying the similarity threshold and measuring the trade-off between precision and recall from which the area under the curve is calculated.\par 
To assess correlation of cosine similarity between latent representation to more interpretable similarities the Tanimoto similarity based on \texttt{RDKit} fingerprints (2048 bits)~\citep{rdkit} and the ROSHAMBO score were computed. The ROSHAMBO score~\citep{atwi2024roshambo} is a shape and colour (pharamcophore) similarity metric for which conformers had to first be aligned in 3D space. \par
\begin{figure}[H]
\begin{center}
\includegraphics[width=0.7\textwidth]{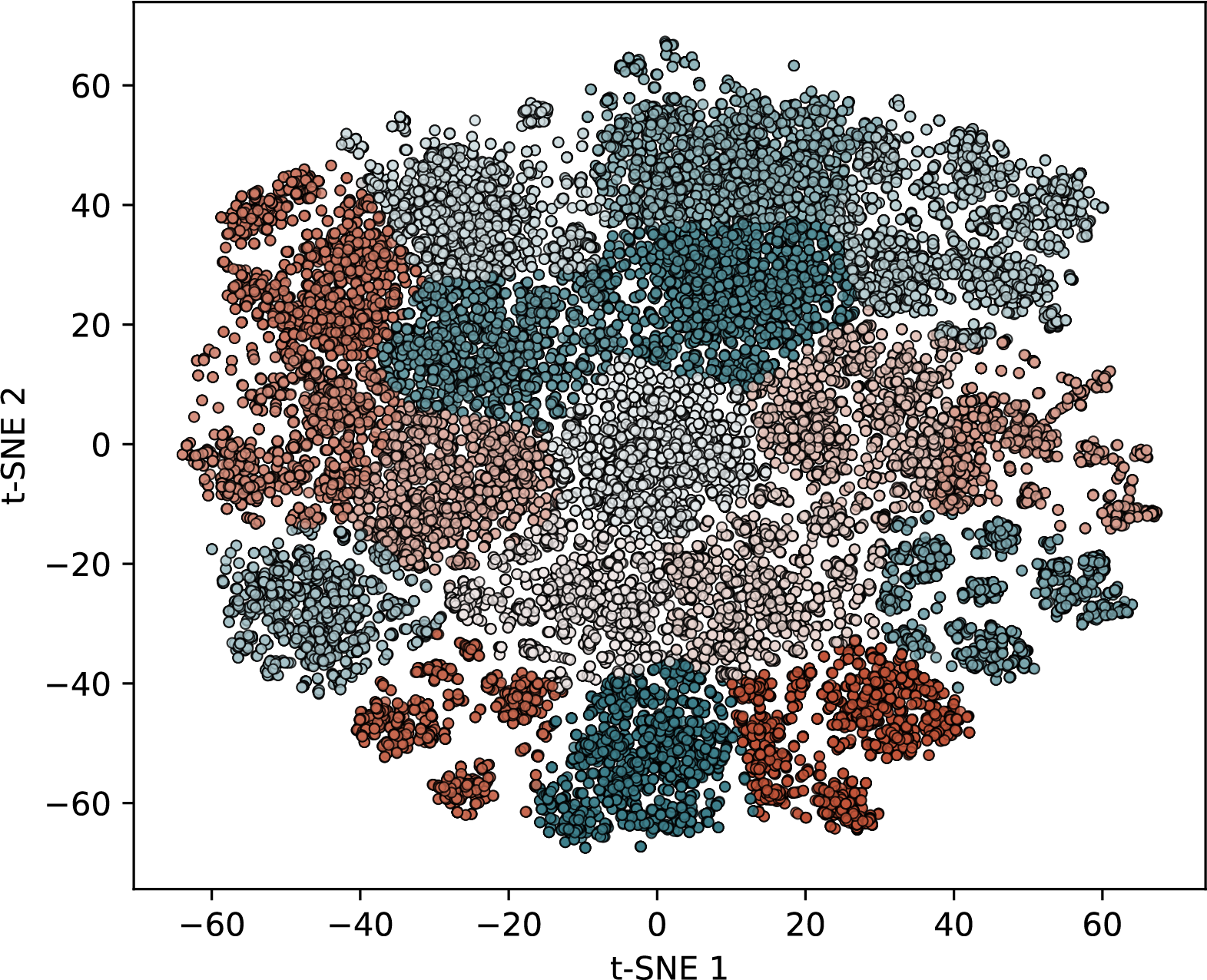}
\end{center}
\caption{t-SNE dimensionality reduction of the fragment library represented as \texttt{RDKit} fingerprints (2048 bits).}
\label{si:fig:tsne_fp}
\end{figure}

The clusters in the t-SNE plots in Figures \ref{fig:enc} and \ref{si:fig:tsne_fp} were obtained using KMeans clustering and the number of clusters was determined by minimizing the silhouette score (29 for latent representation, 18 for fingerprints).

\subsection{Fragment Identification}
\label{si:eval_ident}

The generative pipeline is evaluated on the test set of 100 protein pockets. For each surface 100 samples are generated with number of fragments corresponding to the reference number of fragments. In case of multiple datapoints for one pocket:ligand pair due to graph partitioning a random sample is chosen out of those. Every sample consists of one or more nodes with corresponding centroid and latent representation, which is used to query the library for closest fragment based on cosine similarity. All fragments are subsequently docked using Gnina~\citep{mcnutt2021gnina} as described in Appendix~\ref{si:sampling}. The generative modelling results are compared to the following approaches:
% comparison approaches
\subparagraph{Latent Virtual Screening}
The latent encodings are used directly for virtual screening by ranking the sum of all cosine similarities between all pocket surface points and fragments in the library. The top 100 per target are selected and docked in the full pocket with the bounding box given by the reference ligand.

\subparagraph{Docking Virtual Screening}
We further perform VS with docking. The full library is docked as described above to every test pocket. The top 100 fragments ranked by docking score are selected as hits.

\subparagraph{Random Baseline}
A random baseline is established by randomly selecting fragments out of the library and noising the ground truth centroid by adding the double of a randomly chosen noise level drawn from the normal distribution. For every target the same number of fragments as in the ground truth pocket are sampled.\par

We focus on a few evaluation metrics describing recovery, and proxies for interaction:

\subparagraph{Hard Recovery}
Hard recovery is defined as exact matches between sampled and reference fragments based on non-isomeric SMILES. We ignore stereochemistry as chiral centres might be altered during docking especially with some fragments missing stereo-assignments. We further distinguish between Sampling hits and Unique Fragments. With the first, describing the total number of samples that have a corresponding match while the latter is the unique number of reference fragments that got recovered. Recovery rates are calculated correspondingly by dividing by number of sampled fragments or number of reference fragments that are also in the library. The VS baselines are divided by the total amount of selected top fragments for the Sampling hits (100 fragments x 100 targets).

\subparagraph{Soft Recovery}
Soft recovery metrics are based on the SuCOS score~\citep{leung2019sucos}, which assess shape and colour (pharmacophoric) similarity between a smaller (fragment) and bigger (reference ligand) molecule. We plot the reverse cumulative proportion and assess "soft hits" as those fragments with a score higher than 0.5. The scores are calculated based on docked poses.

\subparagraph{Docking Efficiency}
Docking efficiency distributions are compared in order to assess a proxy for binding affinity. Docking efficiency is the vina score (docking score computed by Gnina~\citep{mcnutt2021gnina}) normalised by the number of heavy atoms to reduce the bias introduced by fragments size. The scores are computed based on docked poses except for the reference, which is based on minimised poses.

\subparagraph{Non-covalent Interactions}
Non-covalent interactions are extracted by PLIP~\citep{plip} applied to all docked fragments individually. Reference profiles are based on crystal structure poses.

%%%%%%%%%%%%%%%%%%%%%%%%%%%%%%%%%%%%%%%%%%%%%%%%%%%%%%%
\section{Additional Results}
\label{si:results}

\subsection{Encoder}

\begin{table}[H]
\caption{Metrics evaluating the encoder on the test set. All scores are computed on the cosine similarity between fragment and protein surface representations with points labelled as true being 3~\si{\angstrom} way from the fragment.}
\label{si:tab:enc_metrics}
\begin{center}
\begin{small}
\begin{tabular}{llcccccccc}
\toprule
Model          & Surface  & \makecell{ROC\\AUC $\uparrow$}    & $\text{EF}_1 \uparrow$ & $\text{EF}_5 \uparrow$ & $\text{EF}_{10} \uparrow$ & \makecell{top-1\\Acc $\uparrow$} &\makecell{top-10\\Acc $\uparrow$} & \makecell{top-100\\Acc $\uparrow$} & \makecell{AUPR $\uparrow$\\(true rate)} \\ \midrule
\multirow{2}{*}{ours}   & whole          & 0.92                   & 22.85                  & 14.35                                      & 8.61                                          & 0.39               & 0.36                                    & 0.33                                     & 0.31 (0.01)                                                                    \\
                        & pocket             & 0.76                  & 2.28                   & 2.23                                       & 2.2                                           & 0.45               & 0.41                                    & 0.39                                     & 0.39 (0.16)                                                                    \\ \midrule
\multirow{2}{*}{no NCI} & whole          & 0.90                                       & 21.37                  & 14.00                                      & 8.47                                          & 0.33               & 0.31                                    & 0.31                                     & 0.28 (0.01)                                                                    \\
                        & pocket                &                0.75                                          & 2.22                   & 2.17                                       & 2.15                                          & 0.41               & 0.39                                    & 0.37                                     & 0.37 (0.16)                                                                  \\ \midrule
\multirow{2}{*}{no FSP} & whole       & 0.92                                 & 24.79                  & 14.29                                      & 8.45                                          & 0.45               & 0.43                                    & 0.36                                     & 0.32 (0.01)                                                                    \\
                        & pocket            & 0.76                 & 3.16                   & 2.70                                       & 2.46                                          & 0.53               & 0.51                                    & 0.42                                     & 0.40 (0.16)                                                                    \\ \bottomrule
\end{tabular}
\end{small}
\end{center}
\end{table}

\begin{table}[H]
\caption{Metrics evaluating the encoder on the validation set. All scores are computed on the cosine similarity between fragment and protein surface representations with points labelled as true being 3~\si{\angstrom} way from the fragment.}
\label{si:tab:enc_metrics_val}
\begin{center}
\begin{small}
\begin{tabular}{llcccccccc}
\toprule
Model          & Surface  & \makecell{ROC\\AUC $\uparrow$}    & $\text{EF}_1 \uparrow$ & $\text{EF}_5 \uparrow$ & $\text{EF}_{10} \uparrow$ & \makecell{top-1\\Acc $\uparrow$} &\makecell{top-10\\Acc $\uparrow$} & \makecell{top-100\\Acc $\uparrow$} & \makecell{AUPR $\uparrow$\\(true rate)} \\ \midrule
\multirow{2}{*}{ours}   & whole     & 0.85    & 17.73    & 11.35      & 7.03      & 0.30        & 0.29   & 0.29   & 0.27 (0.01)   \\
                        & pocket    & 0.72    & 1.91    & 2.04        & 2.07       & 0.34       & 0.34     & 0.35   & 0.36 (0.16)   \\ \midrule
\multirow{2}{*}{no NCI} & whole     & 0.83    & 14.39   & 10.71       & 6.87       & 0.21     & 0.22       & 0.24  & 0.23 (0.01)   \\
                        & pocket   &  0.70    & 1.56    & 1.77        & 1.81      & 0.27     & 0.28       & 0.31    & 0.32 (0.16)   \\ \midrule
\multirow{2}{*}{no FSP} & whole    & 0.86     & 19.44   & 11.25      & 7.00     & 0.34      & 0.33        & 0.31    & 0.28 (0.01)   \\
                        & pocket    & 0.73    & 2.32    & 2.27      & 2.14      & 0.41      & 0.40      & 0.37     & 0.37 (0.16)    \\ \bottomrule
\end{tabular}
\end{small}
\end{center}
\end{table}

\begin{table}[H]
\caption{Average cosine similarities between all fragment embeddings of the fragment library, the training, test and validation set fractions. }
\label{si:tab:frag_sims}
\begin{center}
\begin{small}
\begin{tabular}{lcccc}
\toprule
Model          & avg $h_f$ all $\downarrow$ & avg $h_f$ train $\downarrow$ & avg $h_f$ test $\downarrow$ & avg $h_f$ valid $\downarrow$ \\ \midrule
 \# samples        &  41,224    &   27,463        &11,783   &      5856 \\\midrule
    ours          &     0.25    &    0.25     &    0.28     &      0.26       \\ \midrule
    no NCI        &     0.23    &   0.23      &   0.26      &      0.24      \\ \midrule
    no FSP        &     0.45    &    0.45     &     0.46    &     0.45        \\ \bottomrule
\end{tabular}
\end{small}
\end{center}
\end{table}

\begin{figure}[H]
    \begin{center}
    \includegraphics[width=0.9\linewidth]{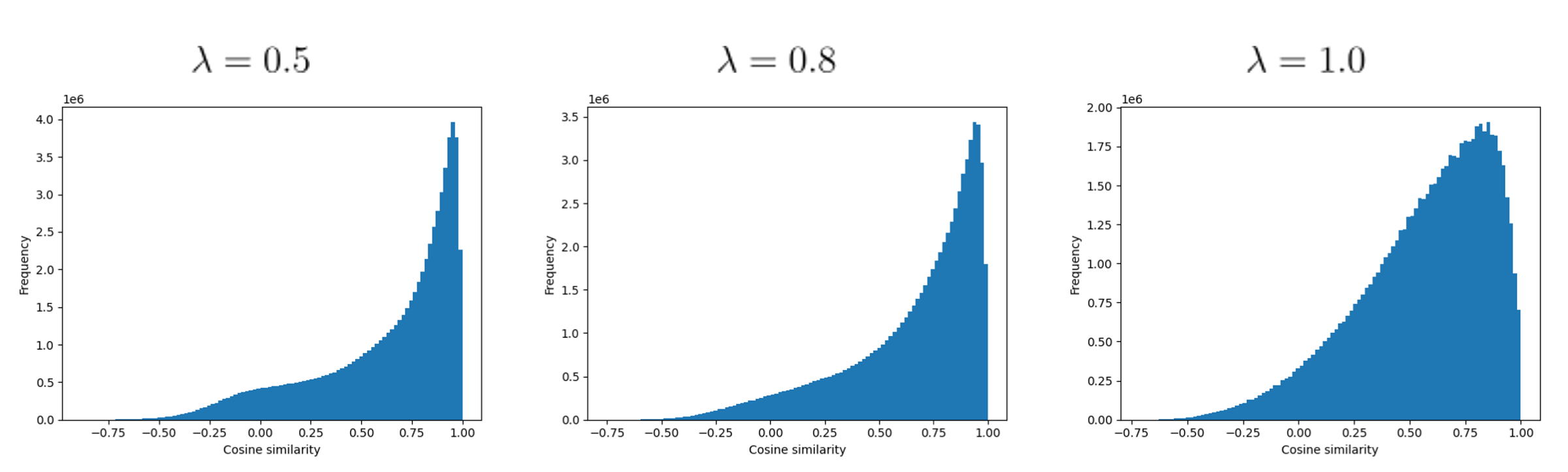}
    \caption{Preliminary results highlighting the importance of the FSP on fragment representation diversity. $\lambda$ corresponds to the loss weight.}
    \label{si:fig:fsp_prelim}
    \end{center}
\end{figure}

\subsection{Fragment Identification}

\begin{figure}[H]
\begin{center}
    \includegraphics[width=0.9\textwidth]{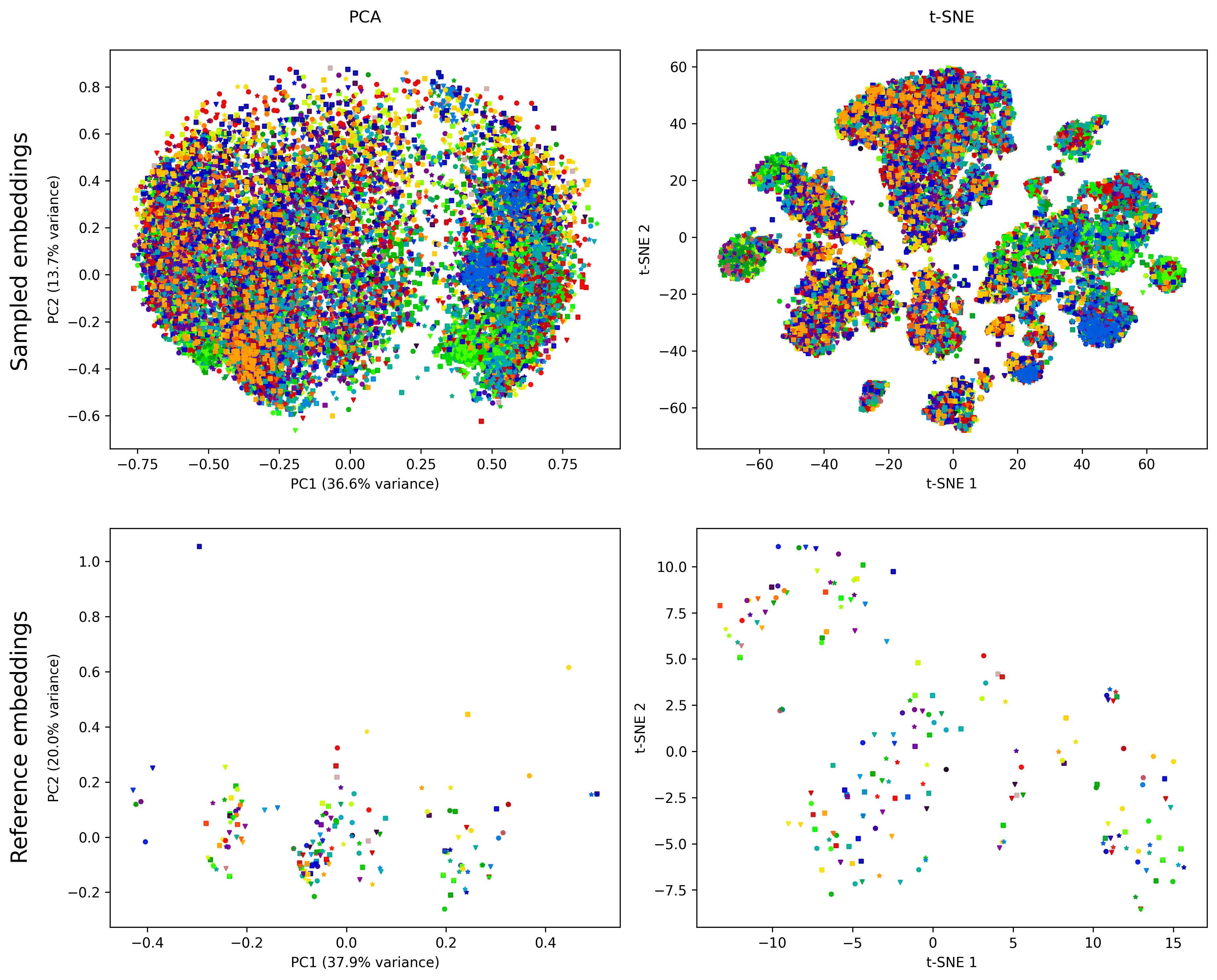}
    \end{center}
    \caption{Dimensionality reduction plot (PCA and t-SNE) of the fragments sampled with LatentFrag (generative pipeline) compared to the reference fragments coloured and symbolised by target.}
    \label{si:fig:pca_target}
\end{figure}

\begin{figure}[H]
\begin{center}
\includegraphics[width=0.5\textwidth]{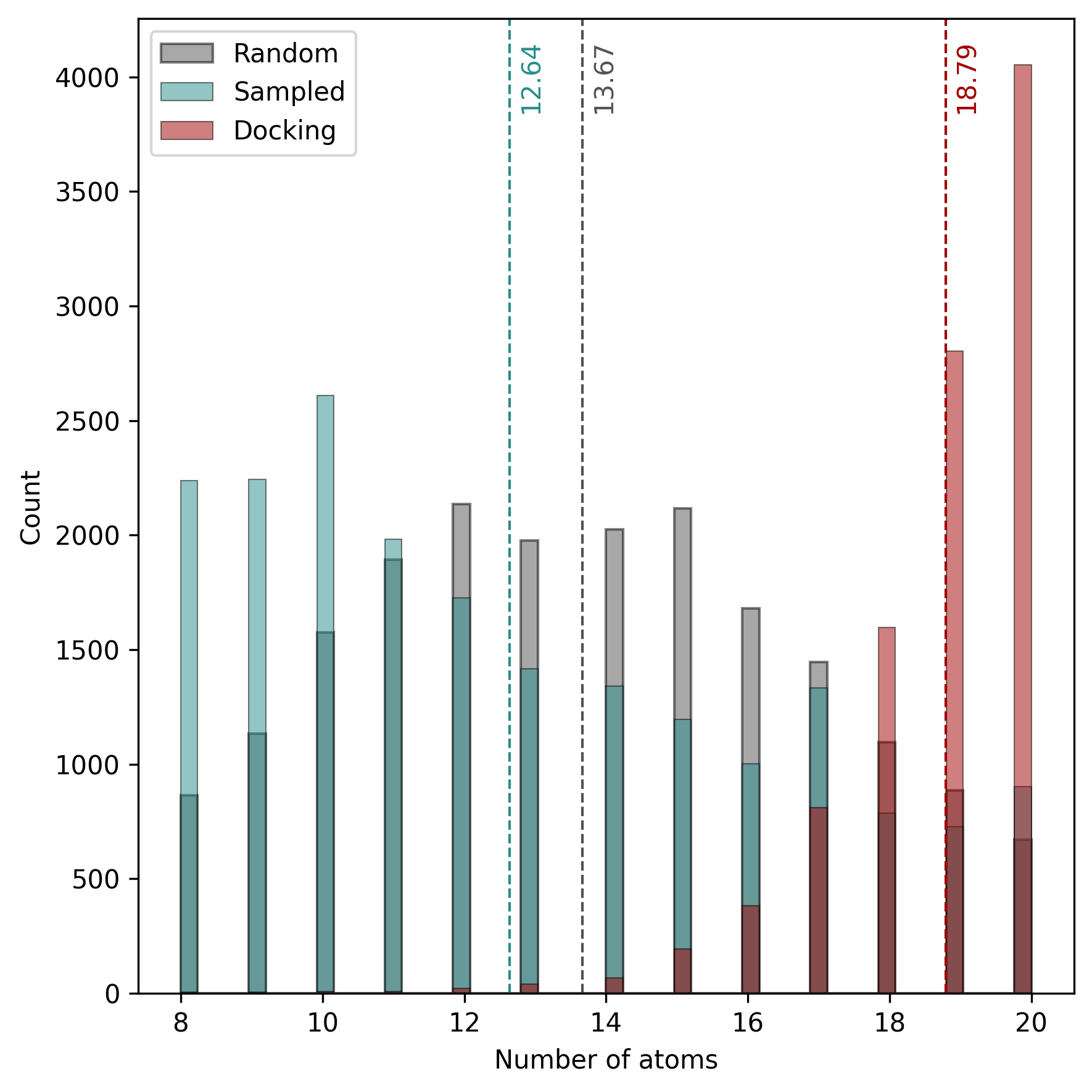}
\end{center}
\caption{Distribution of number of heavy atoms for our LatentFrag generated samples, docking VS and the random baseline.}
\label{si:fig:natoms_hist}
\end{figure}

\begin{table}[ht!]
\caption{Recovery rates in percentage for identical matches to the reference fragments in the test set. Sampling hits refer to how many of the samples are recovering reference fragments while unique fragments is the absolute number of reference fragments that got recovered.}
\label{si:tab:recovery_rates}
\begin{center}
\begin{tabular}{lcccc}
\toprule
                 & LatentFrag & Latent VS & Docking VS & Random \\ \midrule
Sampling hits    & 4.33       & 0.02      & 2.05       & 0.00   \\
Unique fragments & 0.554      & 0.001     & 0.130      & 0.000 \\
\bottomrule
\end{tabular}
\end{center}
\end{table}

\begin{figure}[H]
\begin{center}
\includegraphics[width=0.6\textwidth]{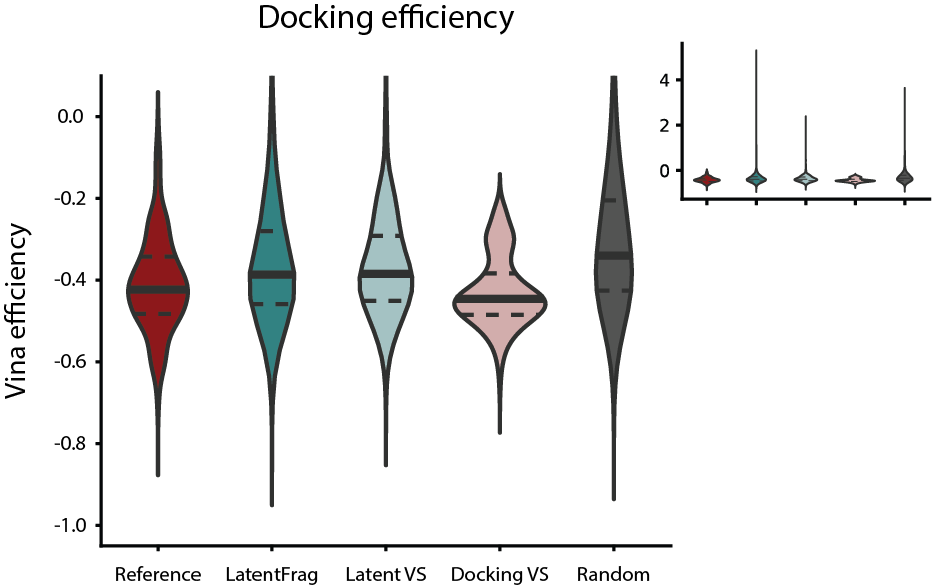}
\end{center}
\caption{Distribution of vina efficiency (vina docking score divided by number of atoms) for the different sampling strategies.}
\label{si:fig:docking_eff_violin}
\end{figure}

\subsubsection{Non-covalent Interactions}

\begin{figure}[H]
	\centering
	\begin{subfigure}[t]{0.6\textwidth}
		\centering
		\includegraphics[width=\textwidth]{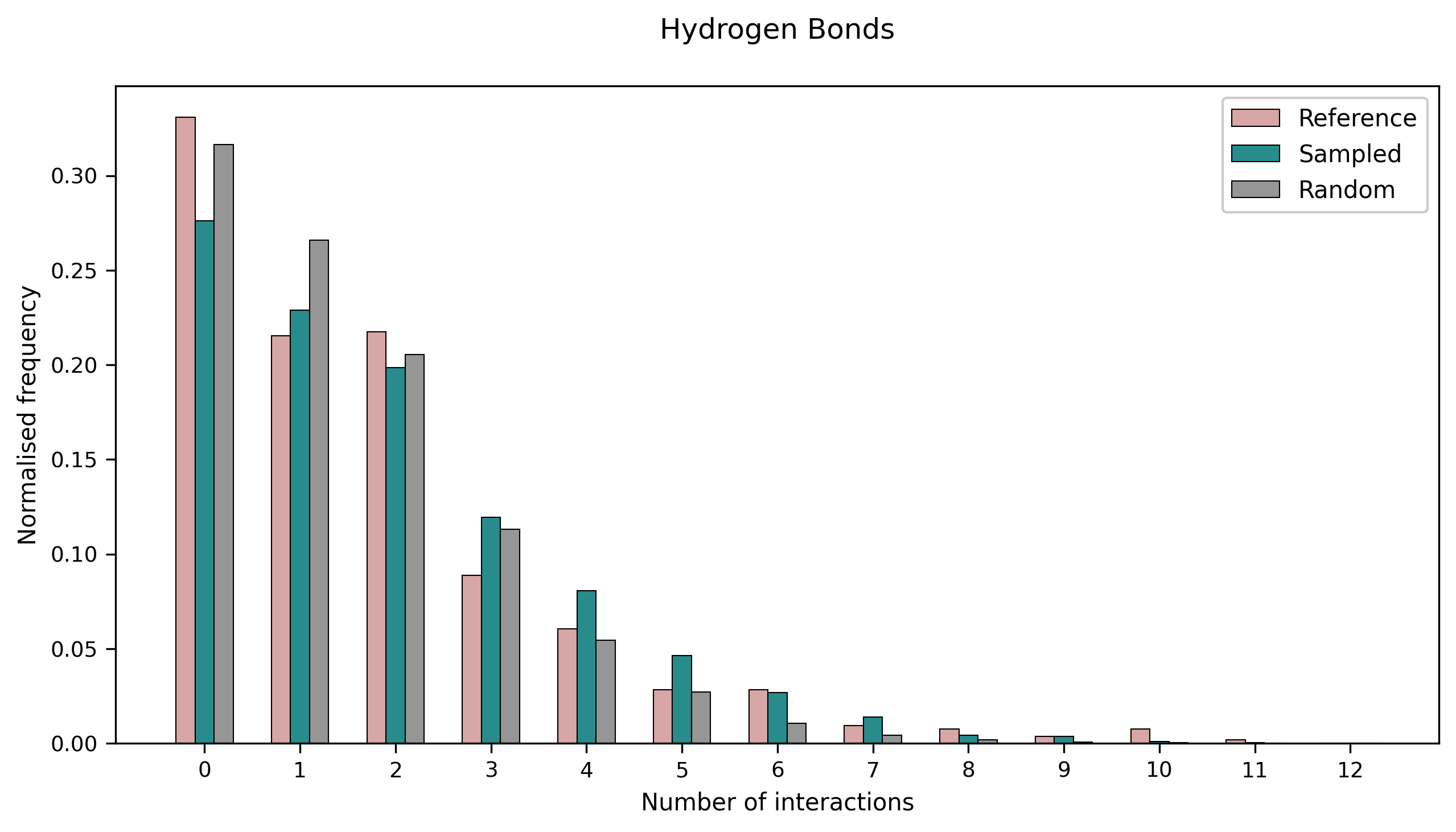}
		\caption{Hydrogen bonds.}
		\label{si:fig:nci:hb}
	\end{subfigure} \hfill
    \begin{subfigure}[t]{0.39\textwidth}
		\centering
		\includegraphics[width=\textwidth]{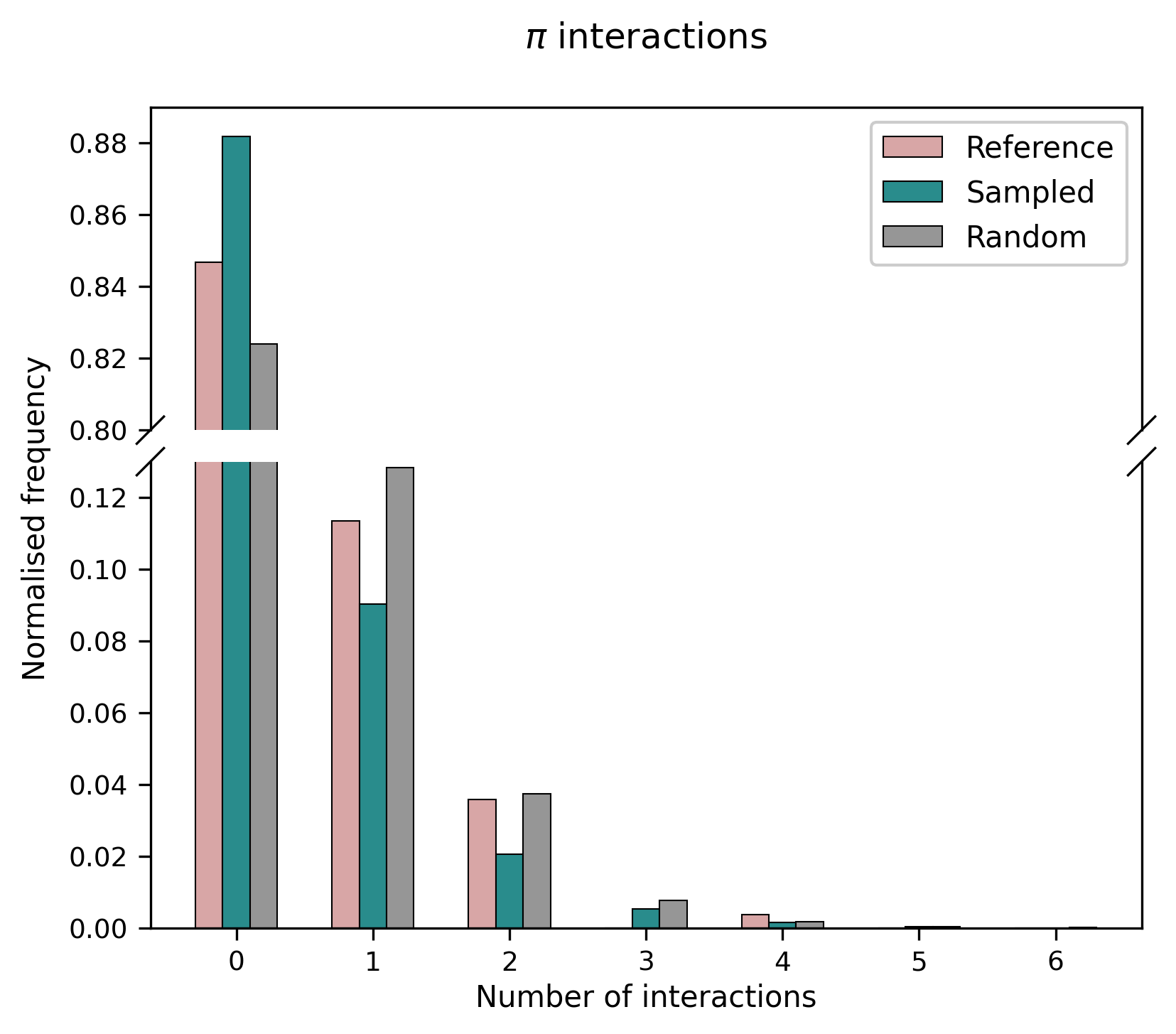}
		\caption{$\pi$ interactions.}
		\label{si:fig:nci:pi}
	\end{subfigure}\hfill
    \centering
	\begin{subfigure}[t]{0.6\textwidth}
		\centering
		\includegraphics[width=\textwidth]{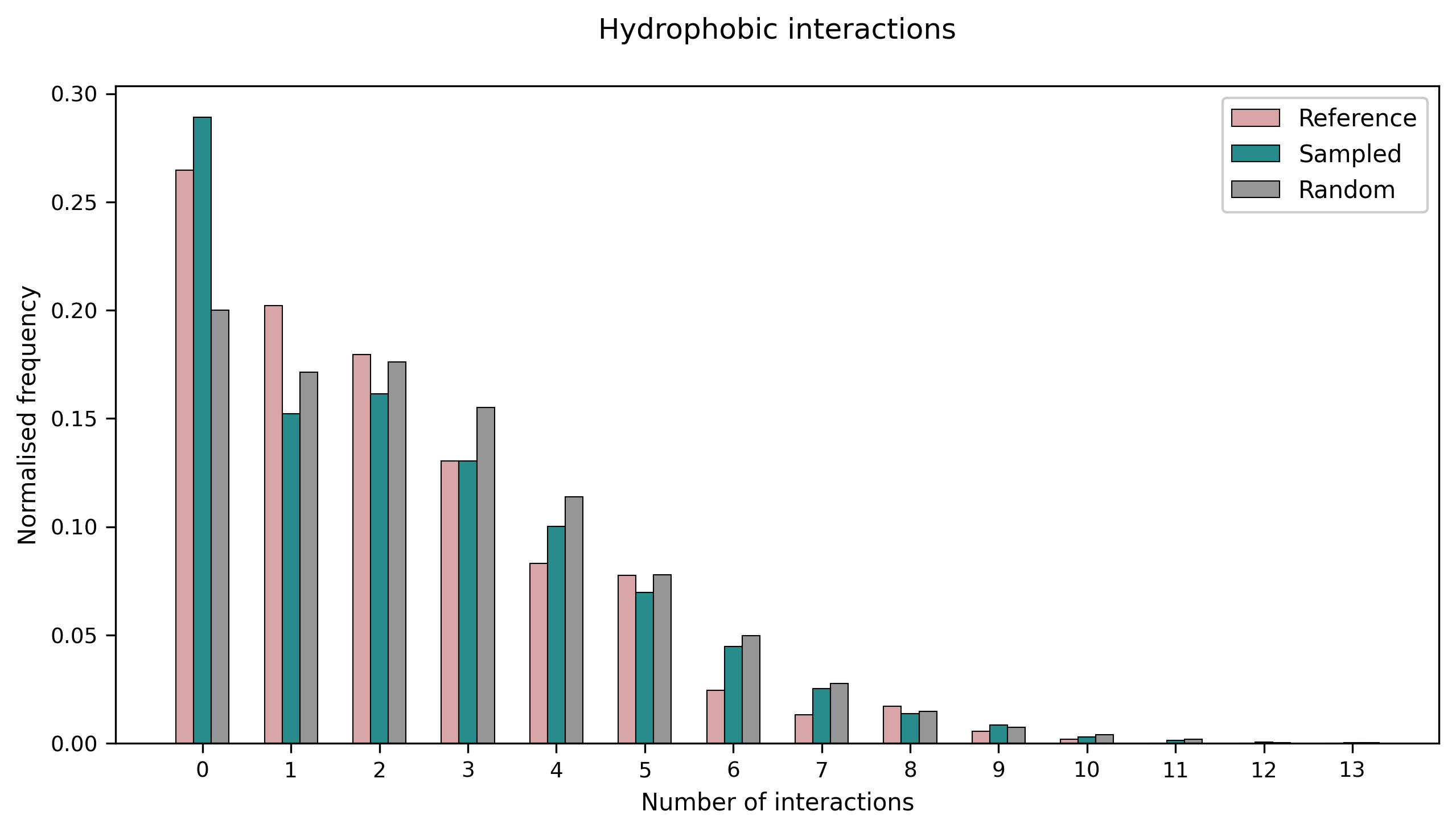}
		\caption{Hydrophobic interactions.}
		\label{si:fig:nci:hydr}
	\end{subfigure} \hfill
    \begin{subfigure}[t]{0.3\textwidth}
		\centering
		\includegraphics[width=\textwidth]{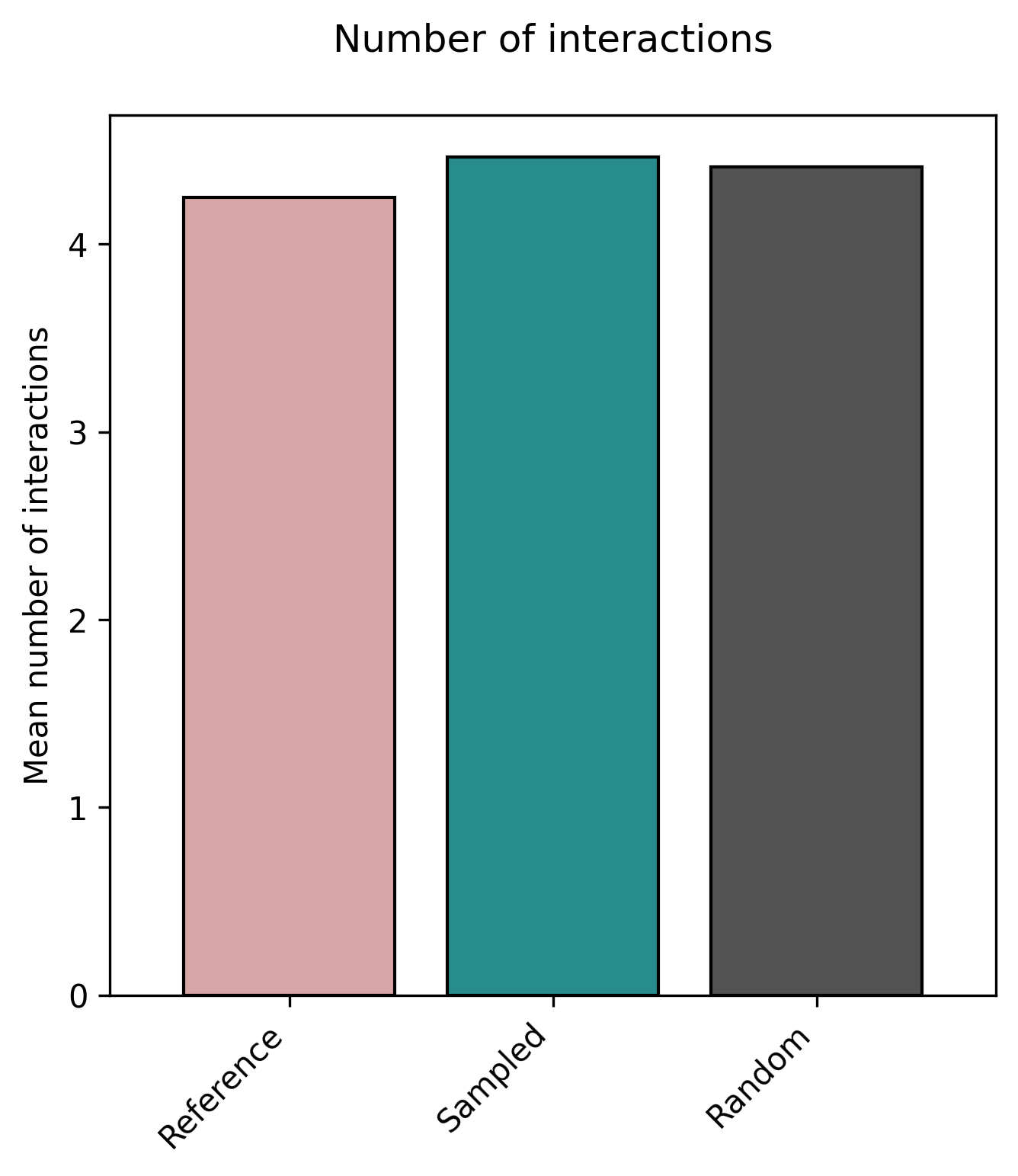}
		\caption{Sum of all interactions.}
		\label{si:fig:nci:total}
	\end{subfigure}
	\caption{Distribution of non-covalent interactions for LatentFrag sampled, reference and random fragments. $\pi$-interactions include $\pi$-$\pi$-stacking and $\pi$-cation interactions. The total number of interactions includes hydrogen bonds, $\pi$-interactions, hydrophobic interactions and salt bridges.}
	\label{si:fig:nci}
\end{figure}

Using the Protein-Ligand Interaction Profiler~(PLIP)~\citep{adasme2021plip} to compare profiles of non-covalent interactions we observed similar interaction type distributions between the reference and samples fragments but a higher average total number of interactions. However, similar results were observed with random samples, with even better profiles than the reference, putting the insightfulness of these results into question. This is likely more an artifact of docking, which aims at maximizing such interactions, than true interactions as we do not dock the reference fragments from crystal structures. Corresponding data is shown in Figure~\ref{si:fig:nci}.

\label{si:sec:nci}

%%%%%%%%%%%%%%%%%%%%%%%%%%%%%%%%%%%%%%%%%%%%%%%%%%%%%%%
\section{Case study}
\label{si:sec:difflinker}
% difflinker approach
We used DiffLinker~\citep{igashov2024equivariant} to generate linkers for two selected fragments sampled with LatentFrag. DiffLinker was provided the two fragments and the same pocket as used for generation to condition the generation on. 50 linkers were generated without providing anchor points or number of atoms to generate to the model. The number of steps for sampling was set to 500.

\begin{figure}[H]
\begin{center}
\includegraphics[width=\textwidth]{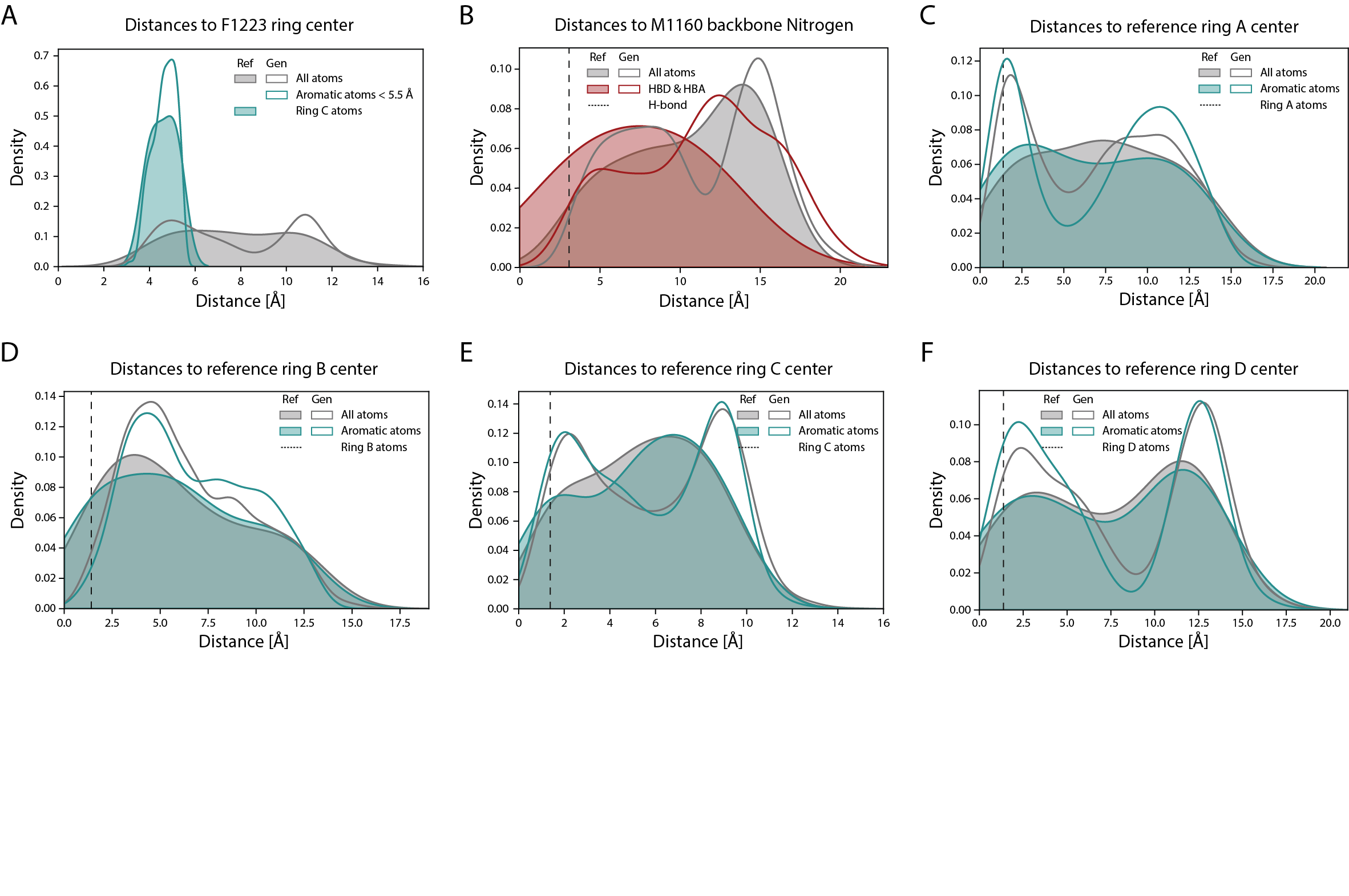}
\end{center}
\caption{
   Kernel density estimates of distance profiles of different atom types (any atom, aromatic atoms, hydrogen bonds donors (HBD) or acceptors (HBD)) in the reference and all generated fragments to different key points in the pocket: \textit{A:} distances to centre of phenylalanine F1223 known to form $\pi-\pi$ interactions with the reference ring of moiety C (filled teal) compared to aromatic atoms of the generated fragments within 5.5~\si{\angstrom} (unfilled teal), \textit{B:} distances to the backbone Nitrogen of methionine M1160 that forms a hydrogen bond with the reference (black dashed line) compared to all HBA/HBA in the reference (filled red) and in the generated fragments (unfilled red), \textit{C-F:} centre of the rings in moieties A, B, C and D of the reference with the average distance to their respective ring atoms (black dashed line) compared to the distances of all aromatic atoms in the reference (filled teal) and the aromatic atoms in the generated fragments (unfilled teal). As comparison, the distributions of all atoms in the reference (filled grey) and all the generated fragments (unfilled grey) are always also shown.
}
\label{si:fig:case_study_kde}
\end{figure}

\begin{figure}[H]
\begin{center}
\includegraphics[width=\textwidth]{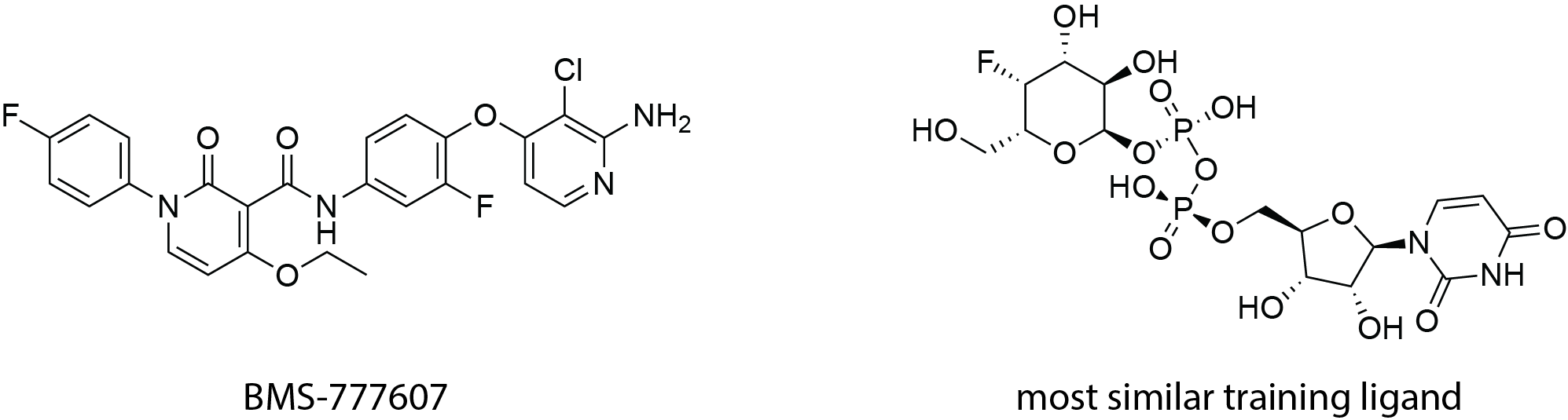}
\end{center}
\caption{Reference ligand BMS-777607 that is bound to c-MET in the crystal structure with PDB ID 6SDD compared to the most similar ligand in our training data. The Tanimoto similarity is 0.75. }
\label{si:fig:case_study_sim_lig}
\end{figure}

\section{Computational Resources}
All models were trained on a single GPU (NVIDIA A100-SXM4-80GB) while development was performed on a NVIDIA GeForce RTX 3090.

\bibliography{ref}

\begin{thebibliography}{58}
\providecommand{\natexlab}[1]{#1}
\providecommand{\url}[1]{\texttt{#1}}
\expandafter\ifx\csname urlstyle\endcsname\relax
  \providecommand{\doi}[1]{doi: #1}\else
  \providecommand{\doi}{doi: \begingroup \urlstyle{rm}\Url}\fi

\bibitem[Adasme et~al.(2021{\natexlab{a}})Adasme, Linnemann, Bolz, Kaiser, Salentin, Haupt, and Schroeder]{adasme2021plip}
Melissa~F Adasme, Katja~L Linnemann, Sarah~Naomi Bolz, Florian Kaiser, Sebastian Salentin, V~Joachim Haupt, and Michael Schroeder.
\newblock Plip 2021: Expanding the scope of the protein--ligand interaction profiler to dna and rna.
\newblock \emph{Nucleic acids research}, 49\penalty0 (W1):\penalty0 W530--W534, 2021{\natexlab{a}}.

\bibitem[Adasme et~al.(2021{\natexlab{b}})Adasme, Linnemann, Bolz, Kaiser, Salentin, Haupt, and Schroeder]{plip}
Melissa~F Adasme, Katja~L Linnemann, Sarah~Naomi Bolz, Florian Kaiser, Sebastian Salentin, V Joachim Haupt, and Michael Schroeder.
\newblock Plip 2021: expanding the scope of the protein–ligand interaction profiler to dna and rna.
\newblock \emph{Nucleic Acids Research}, 49\penalty0 (W1):\penalty0 W530--W534, 05 2021{\natexlab{b}}.
\newblock ISSN 0305-1048.
\newblock \doi{10.1093/nar/gkab294}.
\newblock URL \url{https://doi.org/10.1093/nar/gkab294}.

\bibitem[Albergo \& Vanden-Eijnden(2022)Albergo and Vanden-Eijnden]{albergo2022building}
Michael~S Albergo and Eric Vanden-Eijnden.
\newblock Building normalizing flows with stochastic interpolants.
\newblock \emph{arXiv preprint arXiv:2209.15571}, 2022.

\bibitem[Alimisis et~al.(2021)Alimisis, Davies, Vandereycken, and Alistarh]{spherical_theory}
Foivos Alimisis, Peter Davies, Bart Vandereycken, and Dan Alistarh.
\newblock Distributed principal component analysis with limited communication.
\newblock In M.~Ranzato, A.~Beygelzimer, Y.~Dauphin, P.S. Liang, and J.~Wortman Vaughan (eds.), \emph{Advances in Neural Information Processing Systems}, volume~34, pp.\  2823--2834. Curran Associates, Inc., 2021.
\newblock URL \url{https://proceedings.neurips.cc/paper_files/paper/2021/file/1680e9fa7b4dd5d62ece800239bb53bd-Paper.pdf}.

\bibitem[Atwi et~al.(2024)Atwi, Wang, Sciabola, and Antoszewski]{atwi2024roshambo}
Rasha Atwi, Ye~Wang, Simone Sciabola, and Adam Antoszewski.
\newblock Roshambo: Open-source molecular alignment and 3d similarity scoring.
\newblock \emph{Journal of Chemical Information and Modeling}, 64\penalty0 (21):\penalty0 8098--8104, 2024.

\bibitem[Berman et~al.(2000)Berman, Westbrook, Feng, Gilliland, Bhat, Weissig, Shindyalov, and Bourne]{berman2000protein}
Helen~M Berman, John Westbrook, Zukang Feng, Gary Gilliland, Talapady~N Bhat, Helge Weissig, Ilya~N Shindyalov, and Philip~E Bourne.
\newblock The protein data bank.
\newblock \emph{Nucleic acids research}, 28\penalty0 (1):\penalty0 235--242, 2000.

\bibitem[Bian \& Xie(2018)Bian and Xie]{bian2018computational}
Yuemin Bian and Xiang-Qun Xie.
\newblock Computational fragment-based drug design: current trends, strategies, and applications.
\newblock \emph{The AAPS journal}, 20:\penalty0 1--11, 2018.

\bibitem[Buttenschoen et~al.(2024)Buttenschoen, Morris, and Deane]{buttenschoen2024posebusters}
Martin Buttenschoen, Garrett~M Morris, and Charlotte~M Deane.
\newblock Posebusters: Ai-based docking methods fail to generate physically valid poses or generalise to novel sequences.
\newblock \emph{Chemical Science}, 15\penalty0 (9):\penalty0 3130--3139, 2024.

\bibitem[Chakravarti(2018)]{chakravarti2018distributed}
Suman~K Chakravarti.
\newblock Distributed representation of chemical fragments.
\newblock \emph{Acs Omega}, 3\penalty0 (3):\penalty0 2825--2836, 2018.

\bibitem[Chen et~al.(2023)Chen, ZHANG, and Hinton]{sc}
Ting Chen, Ruixiang ZHANG, and Geoffrey Hinton.
\newblock Analog bits: Generating discrete data using diffusion models with self-conditioning.
\newblock In \emph{The Eleventh International Conference on Learning Representations}, 2023.

\bibitem[Chithrananda et~al.(2020)Chithrananda, Grand, and Ramsundar]{chithrananda2020chemberta}
Seyone Chithrananda, Gabriel Grand, and Bharath Ramsundar.
\newblock Chemberta: large-scale self-supervised pretraining for molecular property prediction.
\newblock \emph{arXiv preprint arXiv:2010.09885}, 2020.

\bibitem[Collie et~al.(2019)Collie, Koh, O’Neill, Stubbs, Khurana, Eddershaw, Snijder, Mauritzson, Barlind, Dale, et~al.]{collie2019structural}
Gavin~W Collie, Cheryl~M Koh, Daniel~J O’Neill, Christopher~J Stubbs, Puneet Khurana, Alice Eddershaw, Arjan Snijder, Fredrik Mauritzson, Louise Barlind, Ian~L Dale, et~al.
\newblock Structural and molecular insight into resistance mechanisms of first generation cmet inhibitors.
\newblock \emph{ACS Medicinal Chemistry Letters}, 10\penalty0 (9):\penalty0 1322--1327, 2019.

\bibitem[Congreve et~al.(2008)Congreve, Chessari, Tisi, and Woodhead]{congreve2008recent}
Miles Congreve, Gianni Chessari, Dominic Tisi, and Andrew~J Woodhead.
\newblock Recent developments in fragment-based drug discovery.
\newblock \emph{Journal of medicinal chemistry}, 51\penalty0 (13):\penalty0 3661--3680, 2008.

\bibitem[Damghani et~al.(2022)Damghani, Elyasi, Pirhadi, Haghighijoo, and Ghazi]{damghani2022type}
Tahereh Damghani, Maryam Elyasi, Somayeh Pirhadi, Zahra Haghighijoo, and Somayeh Ghazi.
\newblock Type ii c-met inhibitors: molecular insight into crucial interactions for effective inhibition.
\newblock \emph{Molecular Diversity}, pp.\  1--13, 2022.

\bibitem[De~Esch et~al.(2021)De~Esch, Erlanson, Jahnke, Johnson, and Walsh]{de2021fragment}
Iwan~JP De~Esch, Daniel~A Erlanson, Wolfgang Jahnke, Christopher~N Johnson, and Louise Walsh.
\newblock Fragment-to-lead medicinal chemistry publications in 2020.
\newblock \emph{Journal of medicinal chemistry}, 65\penalty0 (1):\penalty0 84--99, 2021.

\bibitem[Degen et~al.(2008)Degen, Wegscheid-Gerlach, Zaliani, and Rarey]{degen2008art}
Jorg Degen, Christof Wegscheid-Gerlach, Andrea Zaliani, and Matthias Rarey.
\newblock On the art of compiling and using'drug-like'chemical fragment spaces.
\newblock \emph{ChemMedChem}, 3\penalty0 (10):\penalty0 1503, 2008.

\bibitem[Dwivedi \& Bresson(2020)Dwivedi and Bresson]{dwivedi2020generalization}
Vijay~Prakash Dwivedi and Xavier Bresson.
\newblock A generalization of transformer networks to graphs.
\newblock \emph{arXiv preprint arXiv:2012.09699}, 2020.

\bibitem[Edfeldt et~al.(2011)Edfeldt, Folmer, and Breeze]{edfeldt2011fragment}
Fredrik~NB Edfeldt, Rutger~HA Folmer, and Alexander~L Breeze.
\newblock Fragment screening to predict druggability (ligandability) and lead discovery success.
\newblock \emph{Drug discovery today}, 16\penalty0 (7-8):\penalty0 284--287, 2011.

\bibitem[Enamine()]{enamine-bb}
Enamine.
\newblock Building blocks catalog.
\newblock \url{https://enamine.net/building-blocks/building-blocks-catalog}.
\newblock accessed: 15.09.2025.

\bibitem[Ertl \& Schuffenhauer(2009)Ertl and Schuffenhauer]{ertl2009estimation}
Peter Ertl and Ansgar Schuffenhauer.
\newblock Estimation of synthetic accessibility score of drug-like molecules based on molecular complexity and fragment contributions.
\newblock \emph{Journal of cheminformatics}, 1:\penalty0 1--11, 2009.

\bibitem[Ferla et~al.(2025)Ferla, S{\'a}nchez-Garc{\'\i}a, Skyner, Gahbauer, Taylor, von Delft, Marsden, and Deane]{ferla2025fragmenstein}
Matteo~P Ferla, Rub{\'e}n S{\'a}nchez-Garc{\'\i}a, Rachael~E Skyner, Stefan Gahbauer, Jenny~C Taylor, Frank von Delft, Brian~D Marsden, and Charlotte~M Deane.
\newblock Fragmenstein: predicting protein--ligand structures of compounds derived from known crystallographic fragment hits using a strict conserved-binding--based methodology.
\newblock \emph{Journal of Cheminformatics}, 17\penalty0 (1):\penalty0 4, 2025.

\bibitem[Gainza et~al.(2020)Gainza, Sverrisson, Monti, Rodola, Boscaini, Bronstein, and Correia]{masif}
Pablo Gainza, Freyr Sverrisson, Frederico Monti, Emanuele Rodola, Davide Boscaini, Michael~M Bronstein, and Bruno~E Correia.
\newblock Deciphering interaction fingerprints from protein molecular surfaces using geometric deep learning.
\newblock \emph{Nature Methods}, 17\penalty0 (2):\penalty0 184--192, 2020.

\bibitem[Gao et~al.(2023)Gao, Qiang, Tan, Jia, Ren, Lu, Liu, Ma, and Lan]{gao2023drugclip}
Bowen Gao, Bo~Qiang, Haichuan Tan, Yinjun Jia, Minsi Ren, Minsi Lu, Jingjing Liu, Wei-Ying Ma, and Yanyan Lan.
\newblock Drugclip: Contrastive protein-molecule representation learning for virtual screening.
\newblock \emph{Advances in Neural Information Processing Systems}, 36:\penalty0 44595--44614, 2023.

\bibitem[Guo et~al.(2022)Guo, Guo, Nan, Tian, Iyer, Ma, Wiest, Zhang, Wang, Zhang, et~al.]{guo2022graph}
Zhichun Guo, Kehan Guo, Bozhao Nan, Yijun Tian, Roshni~G Iyer, Yihong Ma, Olaf Wiest, Xiangliang Zhang, Wei Wang, Chuxu Zhang, et~al.
\newblock Graph-based molecular representation learning.
\newblock \emph{arXiv preprint arXiv:2207.04869}, 2022.

\bibitem[Hasselgren \& Oprea(2024)Hasselgren and Oprea]{hasselgren2024artificial}
Catrin Hasselgren and Tudor~I Oprea.
\newblock Artificial intelligence for drug discovery: Are we there yet?
\newblock \emph{Annual Review of Pharmacology and Toxicology}, 64\penalty0 (1):\penalty0 527--550, 2024.

\bibitem[Hu et~al.(2005)Hu, Benson, Smith, Lerner, and Carlson]{hu2005binding}
Liegi Hu, Mark~L Benson, Richard~D Smith, Michael~G Lerner, and Heather~A Carlson.
\newblock Binding moad (mother of all databases).
\newblock \emph{Proteins: Structure, Function, and Bioinformatics}, 60\penalty0 (3):\penalty0 333--340, 2005.

\bibitem[Hubbard \& Murray(2011)Hubbard and Murray]{hubbard2011experiences}
Roderick~E Hubbard and James~B Murray.
\newblock Experiences in fragment-based lead discovery.
\newblock In \emph{Methods in enzymology}, volume 493, pp.\  509--531. Elsevier, 2011.

\bibitem[Igashov et~al.(2022)Igashov, Jamasb, Sadek, Sverrisson, Schneuing, Lio, Blundell, Bronstein, and Correia]{igashov2022decoding}
Ilia Igashov, Arian~R Jamasb, Ahmed Sadek, Freyr Sverrisson, Arne Schneuing, Pietro Lio, Tom~L Blundell, Michael Bronstein, and Bruno Correia.
\newblock Decoding surface fingerprints for protein-ligand interactions.
\newblock \emph{bioRxiv}, pp.\  2022--04, 2022.

\bibitem[Igashov et~al.(2023)Igashov, Schneuing, Segler, Bronstein, and Correia]{retrobridge}
Ilia Igashov, Arne Schneuing, Marwin Segler, Michael Bronstein, and Bruno Correia.
\newblock Retrobridge: Modeling retrosynthesis with markov bridges.
\newblock \emph{arXiv preprint arXiv:2308.16212}, 2023.

\bibitem[Igashov et~al.(2024)Igashov, St{\"a}rk, Vignac, Schneuing, Satorras, Frossard, Welling, Bronstein, and Correia]{igashov2024equivariant}
Ilia Igashov, Hannes St{\"a}rk, Cl{\'e}ment Vignac, Arne Schneuing, Victor~Garcia Satorras, Pascal Frossard, Max Welling, Michael Bronstein, and Bruno Correia.
\newblock Equivariant 3d-conditional diffusion model for molecular linker design.
\newblock \emph{Nature Machine Intelligence}, 6\penalty0 (4):\penalty0 417--427, 2024.

\bibitem[Imrie et~al.(2020)Imrie, Bradley, van~der Schaar, and Deane]{imrie2020deep}
Fergus Imrie, Anthony~R Bradley, Mihaela van~der Schaar, and Charlotte~M Deane.
\newblock Deep generative models for 3d linker design.
\newblock \emph{Journal of chemical information and modeling}, 60\penalty0 (4):\penalty0 1983--1995, 2020.

\bibitem[Jalencas et~al.(2024)Jalencas, Berg, Espeland, Sreeramulu, Kinnen, Richter, Georgiou, Yadrykhinsky, Specker, Jaudzems, et~al.]{jalencas2024design}
Xavier Jalencas, Hannes Berg, Ludvik~Olai Espeland, Sridhar Sreeramulu, Franziska Kinnen, Christian Richter, Charis Georgiou, Vladyslav Yadrykhinsky, Edgar Specker, Kristaps Jaudzems, et~al.
\newblock Design, quality and validation of the eu-openscreen fragment library poised to a high-throughput screening collection.
\newblock \emph{RSC Medicinal Chemistry}, 15\penalty0 (4):\penalty0 1176--1188, 2024.

\bibitem[Jing et~al.(2021)Jing, Eismann, Soni, and Dror]{jing2021equivariant}
Bowen Jing, Stephan Eismann, Pratham~N Soni, and Ron~O Dror.
\newblock Equivariant graph neural networks for 3d macromolecular structure.
\newblock \emph{arXiv preprint arXiv:2106.03843}, 2021.

\bibitem[Leung et~al.(2019)Leung, Bodkin, von Delft, Brennan, and Morris]{leung2019sucos}
Susan Leung, Michael Bodkin, Frank von Delft, Paul Brennan, and Garrett Morris.
\newblock Sucos is better than rmsd for evaluating fragment elaboration and docking poses.
\newblock \emph{ChemRviv preprint 10.26434/chemrxiv.8100203}, 2019.

\bibitem[Li \& Jiang(2021)Li and Jiang]{li2021mol}
Juncai Li and Xiaofei Jiang.
\newblock Mol-bert: An effective molecular representation with bert for molecular property prediction.
\newblock \emph{Wireless Communications and Mobile Computing}, 2021\penalty0 (1):\penalty0 7181815, 2021.

\bibitem[Lipman et~al.(2022)Lipman, Chen, Ben-Hamu, Nickel, and Le]{lipman2022flow}
Yaron Lipman, Ricky~TQ Chen, Heli Ben-Hamu, Maximilian Nickel, and Matt Le.
\newblock Flow matching for generative modeling.
\newblock \emph{arXiv preprint arXiv:2210.02747}, 2022.

\bibitem[Lohmann et~al.(2024)Lohmann, Allenspach, Atz, Schiebroek, Hiss, and Schneider]{lohmann2024protein}
Frederieke Lohmann, Stephan Allenspach, Kenneth Atz, Carl~CG Schiebroek, Jan~A Hiss, and Gisbert Schneider.
\newblock Protein binding site representation in latent space.
\newblock \emph{Molecular Informatics}, pp.\  e202400205, 2024.

\bibitem[Marchand \& Caflisch(2018)Marchand and Caflisch]{SEED}
Jean-Remy Marchand and Amedeo Caflisch.
\newblock In silico fragment-based drug design with seed.
\newblock \emph{EUROPEAN JOURNAL OF MEDICINAL CHEMISTRY}, 156:\penalty0 907--917, AUG 5 2018.
\newblock ISSN 0223-5234.
\newblock \doi{10.1016/j.ejmech.2018.07.042}.

\bibitem[McCorkindale et~al.(2022)McCorkindale, Ahel, Barr, Correy, Fraser, London, Schuller, Shurrush, and Lee]{mccorkindale2022fragment}
William McCorkindale, Ivan Ahel, Haim Barr, Galen~J Correy, James~S Fraser, Nir London, Marion Schuller, Khriesto Shurrush, and Alpha~A Lee.
\newblock Fragment-based hit discovery via unsupervised learning of fragment-protein complexes.
\newblock \emph{bioRxiv}, pp.\  2022--11, 2022.

\bibitem[McNutt et~al.(2021)McNutt, Francoeur, Aggarwal, Masuda, Meli, Ragoza, Sunseri, and Koes]{mcnutt2021gnina}
Andrew~T McNutt, Paul Francoeur, Rishal Aggarwal, Tomohide Masuda, Rocco Meli, Matthew Ragoza, Jocelyn Sunseri, and David~Ryan Koes.
\newblock Gnina 1.0: molecular docking with deep learning.
\newblock \emph{Journal of cheminformatics}, 13\penalty0 (1):\penalty0 43, 2021.

\bibitem[Neeser et~al.(2023)Neeser, Akdel, Kovtun, and Naef]{neeser2023reinforcement}
Rebecca~M Neeser, Mehmet Akdel, Daniel Kovtun, and Luca Naef.
\newblock Reinforcement learning-driven linker design via fast attention-based point cloud alignment.
\newblock \emph{arXiv preprint arXiv:2306.08166}, 2023.

\bibitem[RDKit()]{rdkit}
RDKit.
\newblock {RDK}it: Open-source cheminformatics.
\newblock \url{http://www.rdkit.org}.

\bibitem[Sanner et~al.(1996)Sanner, Olson, and Spehner]{msms}
Michel~F Sanner, Arthur~J Olson, and Jean-Claude Spehner.
\newblock Reduced surface: an efficient way to compute molecular surfaces.
\newblock \emph{Biopolymers}, 38\penalty0 (3):\penalty0 305--320, 1996.

\bibitem[Schneuing et~al.(2024)Schneuing, Igashov, Castiglione, Bronstein, and Correia]{schneuing2024towards}
Arne Schneuing, Ilia Igashov, Thomas Castiglione, Michael~M Bronstein, and Bruno Correia.
\newblock Towards structure-based drug design with protein flexibility.
\newblock In \emph{ICLR 2024 Workshop on Generative and Experimental Perspectives for Biomolecular Design}, 2024.

\bibitem[Schneuing et~al.(2025)Schneuing, Igashov, Dobbelstein, Castiglione, Bronstein, and Correia]{schneuing2025multidomain}
Arne Schneuing, Ilia Igashov, Adrian~W. Dobbelstein, Thomas Castiglione, Michael~M. Bronstein, and Bruno Correia.
\newblock Multi-domain distribution learning for de novo drug design.
\newblock In \emph{The Thirteenth International Conference on Learning Representations}, 2025.
\newblock URL \url{https://openreview.net/forum?id=g3VCIM94ke}.

\bibitem[scitkit learn()]{curv_classifier}
scitkit learn.
\newblock Histgradientboostingclassifier.
\newblock \url{https://scikit-learn.org/stable/modules/generated/sklearn.ensemble.HistGradientBoostingClassifier.html}.

\bibitem[Shin et~al.(2019)Shin, Park, Kang, and Ho]{shin2019self}
Bonggun Shin, Sungsoo Park, Keunsoo Kang, and Joyce~C Ho.
\newblock Self-attention based molecule representation for predicting drug-target interaction.
\newblock In \emph{Machine learning for healthcare conference}, pp.\  230--248. PMLR, 2019.

\bibitem[Shoemake(1985)]{slerp}
Ken Shoemake.
\newblock Animating rotation with quaternion curves.
\newblock \emph{SIGGRAPH Comput. Graph.}, 19\penalty0 (3):\penalty0 245–254, July 1985.
\newblock ISSN 0097-8930.
\newblock \doi{10.1145/325165.325242}.
\newblock URL \url{https://doi.org/10.1145/325165.325242}.

\bibitem[Somnath et~al.(2021)Somnath, Bunne, and Krause]{somnath2021multi}
Vignesh~Ram Somnath, Charlotte Bunne, and Andreas Krause.
\newblock Multi-scale representation learning on proteins.
\newblock \emph{Advances in Neural Information Processing Systems}, 34:\penalty0 25244--25255, 2021.

\bibitem[Sverrisson et~al.(2021)Sverrisson, Feydy, Correia, and Bronstein]{dmasif}
Freyr Sverrisson, Jean Feydy, Bruno~E Correia, and Michael~M Bronstein.
\newblock Fast end-to-end learning on protein surfaces.
\newblock In \emph{Proceedings of the IEEE/CVF Conference on Computer Vision and Pattern Recognition}, pp.\  15272--15281, 2021.

\bibitem[Tong et~al.(2023)Tong, Fatras, Malkin, Huguet, Zhang, Rector-Brooks, Wolf, and Bengio]{tong2023improving}
Alexander Tong, Kilian Fatras, Nikolay Malkin, Guillaume Huguet, Yanlei Zhang, Jarrid Rector-Brooks, Guy Wolf, and Yoshua Bengio.
\newblock Improving and generalizing flow-based generative models with minibatch optimal transport.
\newblock \emph{arXiv preprint arXiv:2302.00482}, 2023.

\bibitem[Vignac et~al.(2022)Vignac, Krawczuk, Siraudin, Wang, Cevher, and Frossard]{vignac2022digress}
Clement Vignac, Igor Krawczuk, Antoine Siraudin, Bohan Wang, Volkan Cevher, and Pascal Frossard.
\newblock Digress: Discrete denoising diffusion for graph generation.
\newblock \emph{arXiv preprint arXiv:2209.14734}, 2022.

\bibitem[Wang et~al.(2021)Wang, Li, Jin, Cho, Ji, Han, and Burke]{wang2021chemical}
Hongwei Wang, Weijiang Li, Xiaomeng Jin, Kyunghyun Cho, Heng Ji, Jiawei Han, and Martin~D Burke.
\newblock Chemical-reaction-aware molecule representation learning.
\newblock \emph{arXiv preprint arXiv:2109.09888}, 2021.

\bibitem[Word et~al.(1999)Word, Lovell, Richardson, and Richardson]{word1999asparagine}
J~Michael Word, Simon~C Lovell, Jane~S Richardson, and David~C Richardson.
\newblock Asparagine and glutamine: using hydrogen atom contacts in the choice of side-chain amide orientation.
\newblock \emph{Journal of molecular biology}, 285\penalty0 (4):\penalty0 1735--1747, 1999.

\bibitem[Yang et~al.(2024)Yang, Xiang, and Li]{yang20243d}
Bo~Yang, Chijian Xiang, and Jianing Li.
\newblock 3d structure-based generative small molecule drug design: Are we there yet?
\newblock \emph{bioRxiv}, pp.\  2024--12, 2024.

\bibitem[Yu et~al.(2020)Yu, Modugula, Ichihara, Kramschuster, Keng, Abel, and Wang]{yu2020general}
Haoyu~S Yu, Kalyan Modugula, Osamu Ichihara, Kimberly Kramschuster, Simon Keng, Robert Abel, and Lingle Wang.
\newblock General theory of fragment linking in molecular design: why fragment linking rarely succeeds and how to improve outcomes.
\newblock \emph{Journal of Chemical Theory and Computation}, 17\penalty0 (1):\penalty0 450--462, 2020.

\bibitem[Zhang et~al.(2014)Zhang, Ai, Shi, Peng, Ji, Liu, Geng, and Li]{zhang2014discovery}
Wei Zhang, Jing Ai, Dakuo Shi, Xia Peng, Yinchun Ji, Jian Liu, Meiyu Geng, and Yingxia Li.
\newblock Discovery of novel type ii c-met inhibitors based on bms-777607.
\newblock \emph{European Journal of Medicinal Chemistry}, 80:\penalty0 254--266, 2014.

\bibitem[Zhou et~al.(2023)Zhou, Gao, Ding, Zheng, Xu, Wei, Zhang, and Ke]{zhouuni}
Gengmo Zhou, Zhifeng Gao, Qiankun Ding, Hang Zheng, Hongteng Xu, Zhewei Wei, Linfeng Zhang, and Guolin Ke.
\newblock Uni-mol: A universal 3d molecular representation learning framework.
\newblock In \emph{The Eleventh International Conference on Learning Representations}, 2023.

\end{thebibliography}

\end{document}